\documentclass[%
 reprint,
 amsmath,amssymb,
 aps,
prb,
]{revtex4-2}

\usepackage{graphicx, color}
\usepackage{dcolumn}
\usepackage{bm}
\usepackage{hyperref}
\hypersetup{
 citecolor=blue,
 colorlinks = true,
 urlcolor = blue
}
\usepackage[normalem]{ulem}
\usepackage{physics}
\allowdisplaybreaks[1]

\begin{document}

\title{Electrical conductivity and screening effect of spin-1 chiral fermions scattered by charged impurities}

\author{Risako Kikuchi}
\affiliation{Department of Physics, Nagoya University, Nagoya 464-8602, Japan}
\author{Ai Yamakage}
\affiliation{Department of Physics, Nagoya University, Nagoya 464-8602, Japan}

\date{\today}

\begin{abstract}
We theoretically study the quantum transport in a three-dimensional spin-1 chiral fermion system in the presence of coulomb impurities based on the self-consistent Born approximation. 
We find that the flat-band states anomalously enhance the screening effect, and the electrical conductivity is increased in the low-energy region. 
It is also found that reducing the screening length leads to an increase in the forward scattering contribution and, thus, an increase in the vertex correction in the high-energy region.
\end{abstract}

\maketitle

\section{\label{sec:level1}Introduction}

In chiral crystals, energy bands can host topologically protected multifold degeneracies \cite{Ma2012, Beyond2016}. 
They are regarded as multifold chiral fermions; twofold fermions are simply Weyl (or Dirac) fermions, while threefold, sixfold, and eightfold fermions have no counterpart in the standard theory. 
Therefore, the multifold fermions have the potential to exhibit unique quantum phenomena. 
We focus on the threefold fermion, which hosts two linear bands and one momentum-independent flat band around the threefold degenerate point, called ``spin-1 chiral fermion" or ``triple-component fermion (TCF)."
The three-dimensional spin-1 fermion have been theoretically predicted to appear in chiral crystals
\cite{Tang2017-kk, Chang2017} and have been observed in CoSi \cite{Pshenay-Severin_2018, Takane2019, Rao2019-ts}, RhSi \cite{Sanchez2019-by, Mozaffari2020}, RhSn \cite{Li2019_RhSn}, and in the sixfold fermion in AlPt \cite{Schr2019}, which is topologically equivalent to two copies of spin-1 fermions.
In recent years, quantum transport of the spin-1 fermion \cite{Wu_2019, Xu2019, Dutta2021, Tang21, Kikuchi2022, Petrova2023}, optical response \cite{Flicker18, Sanchez-Martinez2019-ek, Li2019, Habe2019-ss, Maulana2020, Dylan2020, Xu2020-zb, Le2020, Ni2020-ze, Ni2021-eo, Rees2021, Kaushik2021-hx, Dey2022-th, Lu2022}, thermal conduction phenomena \cite{Pshenay-Severin_2018_thermo, Sk_2022}, quadratic dispersion with the spin-1 structure \cite{Nandy2019-qw, Chen2021-dv, Pal2022-aw}, and quantum phenomena originating from topological structures \cite{Yuan2019-hz, Huber2022, hsu2022disorder} have been studied, revealing exotic quantum phenomena.
A two-dimensional (approximate) version of spin-1 chiral fermion has also been studied to show a peculiar quantum transport phenomenon originating from the flat band \cite{Bercioux2009, Shen2010, Vigh2013, Hausler2015-lp, Yang2019-oj, Burgos2022-rl}.

We previously showed the quantum transport of a three-dimensional spin-1 fermion under the Gaussian impurity potential and have found a peak of the density of states (DOS) and the suppressed electrical conductivity near zero Fermi energy \cite{Kikuchi2022}. 
This peculiar behavior is attributed to the interband effect between the flat and linear bands. 
On the other hand, in that study, the screening length was assumed to be a parameter. 
In reality, the screening length is not a parameter but depends on the DOS, which could be strongly modified by the impurity potential. 
Furthermore, the type of impurity potential may alter the transport property. 
Weyl fermion, for example, undergoes a metal-semimetal transition at the band degeneracy point under the Gaussian impurity but not the Coulomb impurity \cite{0minato2014, Kobayashi2014, Nandkishore2014-vz, Ominato2015-um, Ominato2016-jl}.

In this study, we investigate quantum transport phenomena in the spin-1 fermion systems subject to impurity scattering due to the Coulomb potential. 
For the analysis, we use self-consistent Born approximation (SCBA) in the linear response theory to correctly incorporate the screening effect associated with the broadening of the spectral function. 
Our results indicate that the screening effect is strongly enhanced around the zero energy as the number of flat-band states included in the theory increases.
This screening effect give rise to a peak of the electrical conductivity around the zero energy. 
We also find a substantial vertex correction effect impacting conductivity in the high-energy region, where the screening effect diminishes.

The paper is organized as follows. 
Our model for a spin-1 chiral fermion in the presence of impurity is introduced in Sec.~\ref{model}.
We calculate the DOS and the conductivity within the SCBA with the current vertex correction in Sec.~\ref{linear1}. 
The self-consistent equation for self-energy and the Bethe-Salpeter equation for vertex correction are explicitly shown.
Our findings are detailed in Sec.~\ref{linear2}.  
The conductivity within the Boltzmann theory is calculated in Sec.~\ref{Boltzmann}, followed by a discussion of the roles of the interband effect and vertex correction in Sec.~\ref{discussion}. 
We also discuss the validity of the approximation and potential experimental implications.
Section \ref{conclusion} summarizes this work.

\section{model}
\label{model}
First, we introduce a model for a spin-1 fermion, Coulomb-impurity potential, and screening length to calculate the electrical conductivity. 
The Hamiltonian of a three-dimensional spin-1 fermion is written as
\begin{align}
\hat{\mathcal{H}}=\hbar v\hat{\bm{S}}\cdot \bm{k},
\end{align}
where $v$ is the Fermi velocity, $\bm k$ is the electron wavenumber, and $\bm{S}=(\hat{S}_{x},\hat{S}_{y},\hat{S}_{z})$ are the spin operators: 
\begin{align}
\hat{S}_{x} &=
\pmqty{
0 & i & 0\\
-i & 0 & 0\\
0 & 0 & 0
},
\\
\hat{S}_{y}&=
\pmqty{
0 & 0 & -i\\
0 & 0 & 0\\
i & 0 & 0
},
 \\
\hat{S}_{z} &=
\begin{pmatrix}
0 & 0 & 0\\
0 & 0 & i\\
0 & -i & 0\\
\end{pmatrix}.
\end{align}
From the Hamiltonian, the energy eigenvalues are obtained as
\begin{align}
 \epsilon_{\lambda,\bm{k}}=\hbar v\lambda k,
\end{align}
where $\lambda$ is the label for the conduction band ($\lambda = 1$), the flat band ($\lambda = 0$), and the valence band ($\lambda = -1$).
The eigenstates $\vb*{v}_{\lambda, \vb*{k}}$ of the spin-1 fermion system are written as
\begin{align}
 \vb*{v}_{+1, \vb*{k}} 
&=\frac{1}{\sqrt{2}k\sqrt{k_x^2+k_z^2}}\left( \begin{array}{c} k_yk_z-ikk_x \\ -k_x^2-k_z^2 \\ k_xk_y+ikk_z \end{array} \right),\label{eigenstates1}\\
\vb*{v}_{0, \vb*{k}}
&=\frac{1}{k}\left( \begin{array}{c} k_z \\ k_y \\ k_x \end{array} \right),\\
\vb*{v}_{-1, \vb*{k}}
&=\frac{1}{\sqrt{2}k\sqrt{k_x^2+k_z^2}}\left( \begin{array}{c} k_yk_z+ikk_x \\ -k_x^2-k_z^2 \\ k_xk_y-ikk_z \end{array} \right).\label{eigenstates2}
\end{align}

We assume the screened Coulomb potential as impurity potential defined by
\begin{align}
U(\bm{r}) &=  \pm \frac{e^2}{\kappa r}\exp(-q_{\text{s}} r).
\label{the Coulomb}
\end{align}
where $\kappa$ denotes the static dielectric constant, and the double sign $\pm$ assumes an equal number of positive and negative charged impurities, ensuring that the Fermi level is fixed irrelevant to the impurity concentration. 
The Thomas-Fermi screening length $q_{\text{s}}^{-1}$ is given by
\begin{align}
q_{\text{s}}^2 &= \frac{4\pi e^2}{\kappa}D(\epsilon),
\label{Thomas-Fermi}
\end{align}
at zero temperature where $D(\epsilon)$ is the DOS.

We define a parameter characterizing the scattering strength, which is an effective fine-structure constant, as
\begin{align}
\alpha &= \frac{e^2}{\hbar v\kappa}.
\end{align}
The Fourier transform of Eq.~(\ref{the Coulomb}) is obtained to be
\begin{align}
u(\bm{q}) &= 
 \int d \vb*{r} e^{-i \vb*{q} \cdot \vb*{r}} U(\vb*{r})
 \notag\\&
 =
 \pm
\frac{4\pi e^2}{\kappa (q^2+q_{\text{s}}^2)}\label{u_the Coulomb}
\notag\\
&=\pm\frac{1}{(q^2/4\pi\hbar v\alpha)+D(\epsilon)}.
\end{align}
The isotropic disorder potential is characterized by the moment of scattering angle as
\begin{align}
V_n^2(k,k') = 2\pi\int_{-1}^1 d(\cos\theta_{\bm{kk'}})|u(\bm{k}-\bm{k'})|^2\cos^n\theta_{\bm{kk'}},
\label{V}
\end{align}
where $\theta_{\bm{kk'}}$ represents the angle between $\bm{k}$ and $\bm{k'}$, and explicitly shown as
\begin{align}
V_0^2(k,k')
&= \frac{e^4\pi^3}{\kappa^2}\frac{64}{C(k,k')^2-4k^2k'^2}, \\
V_1^2(k,k')
&= \frac{e^4\pi^3}{\kappa^2}\frac{16}{k^2k'^2}\left(\frac{2C(k,k')kk'}{C(k,k')^2-4k^2k'^2}
\right.\notag\\ &\left.\quad
-\text{arctanh} \left(\frac{2kk'}{C(k,k')}\right)\right), \\
V_2^2(k,k')
&= \frac{e^4\pi^3}{\kappa^2}\frac{16}{k^3k'^3}\left(\frac{2C(k,k')^2kk'-4k^3 k'^3}{C(k,k')^2-4k^2k'^2}
\right.\notag\\ &\left.\quad
-C(k,k')\text{arctanh} \left(\frac{2kk'}{C(k,k')}\right)\right), \\
V_3^2(k,k')
&= \frac{e^4\pi^3}{\kappa^2}\frac{4C(k,k')}{k^4k'^4}\left(\frac{6C(k,k')^2kk'-16k^3 k'^3}{C(k,k')^2-4k^2k'^2}
\right.\notag\\ &\left.\quad
-3C(k,k')\text{arctanh} \left(\frac{2kk'}{C(k,k')}\right)\right),
\end{align}
where
\begin{align}
C(k,k') \equiv k^2+k'^2+q_s^2.
\end{align}
We also define 
\begin{align}
q_0 &= n_{\mathrm i}^{{1}/{3}}\label{coulombq_0},
\end{align}
where $n_{\text{i}}$ is the number of scatterers per unit volume.

\section{\label{linear1}The linear response theory (SCBA)-Formulation}
Next, we calculate the DOS and conductivity by SCBA in the linear response theory.

\subsection{Formulation}

Assuming a uniformly random impurity distribution, the impurity-averaged Green function is expressed as
\begin{align}
\hat{G}(\bm{k},\epsilon +is0) 
&=& \frac{1}{\epsilon \hat{S}_0-\hbar vk\hat{\bm{S}}\cdot\bm{n}-\hat{\Sigma}(\bm{k},\epsilon+is0)}, 
\label{green function}
\end{align}
where $\hat{S}_0$ is the identity matrix, $\bm{n}=\bm{k}/k$ is the unit vector and the sign $s$ means the retarded ($s=1$) and advanced ($s=-1$) Green's functions.
The self-energy is written as
\begin{align}
\hat{\Sigma}(\bm{k},\epsilon+is0)
 = \int\frac{d\bm{k'}}{(2\pi)^3}n_{\text{i}}|u(\bm{k}-\bm{k'})|^2\hat{G}(\bm{k'},\epsilon+is0).
 \label{self energy}
\end{align}
by SCBA.
The DOS per unit volume is written as
\begin{align}
D(\epsilon) = -\frac{1}{\pi}\Im\int\frac{d\bm{k}}{(2\pi)^3}\Tr\hat{G}(\bm{k},\epsilon+i0).
\label{dos}
\end{align}
The conductivity is calculated as
\begin{align}
\sigma(\epsilon) &= -\frac{\hbar e^2 v^2}{4\pi}\sum _{s,s'=\pm1}ss'\int\frac{d\bm{k'}}{(2\pi)^3}\text{Tr} [\hat{S}_x\hat{G}(\bm{k'},\epsilon+is0)\nonumber\\
& \quad
\times\hat{J}_x(\bm{k'},\epsilon+is0,\epsilon+is'0)\hat{G}(\bm{k'},\epsilon+is'0)]\label{conductivity}
\end{align}
by the Kubo formula.
$\hat{J}_x(\bm{k},\epsilon,\epsilon')$ is the current vertex part in the $x$ direction and is determined by the Bethe-Salpeter equation:
\begin{align}
\hat{J}_x(\bm{k},\epsilon,\epsilon') &=
\hat{S}_x + \int\frac{d\bm{k'}}{(2\pi)^3}n_{\text{i}}|u(\bm{k}-\bm{k'})|^2\hat{G}(\bm{k'},\epsilon)\nonumber\\
&\quad\times
\hat{J}_x(\bm{k'},\epsilon,\epsilon')\hat{G}(\bm{k'},\epsilon').\label{Bethe}
\end{align}
As demonstrated in the previous study \cite{Kikuchi2022}, the equation mentioned above can be reformulated as an $8 \times 8$ matrix equation, which is conveniently reproduced in Appendix~\ref{j} for the reader's benefit.

\subsection{Intraband and interband contributions}

For a comprehensive understanding of physical quantities in multi-orbital systems, it is advantageous to decompose them into intraband and interband components.
They are obtained by diagonalizing the Green's function matrix as
\begin{align}
\hat{U}^{\dagger}\hat{G}(\bm{k},\epsilon+is0)\hat{U}=
\begin{pmatrix}
G^s_c & 0 & 0\\
0 & G^s_0 & 0\\
0 & 0 & G^s_v\\
\end{pmatrix}.
\end{align}
where the subscripts $c$, $0$, and $v$ denote the conduction, flat, and valence bands in the band basis, respectively.
In this basis, the velocities $S_x$ and $J_x$ are written as
\begin{align}
\hat{U}^{\dagger}\hat{S}_{x}\hat{U}=
\begin{pmatrix}
S_{cc} & S_{c0} & 0\\
S_{0c} & 0 & S_{0v}\\
0 & S_{v0} & S_{vv}\\
\end{pmatrix},
\end{align}
and
\begin{align}
\hat{U}^{\dagger}\hat{J_{x}}(k, \epsilon+is0, \epsilon+is'0)\hat{U}=
\begin{pmatrix}
J^{ss'}_{cc} & J^{ss'}_{c0} & 0\\
J^{ss'}_{0c} & J^{ss'}_{00} & J^{ss'}_{0v}\\
0 & J^{ss'}_{v0} & J^{ss'}_{vv}\\
\end{pmatrix}.
\end{align}
We decompose the DOS into the Dirac-cone and the flat-band terms as
\begin{align}
D_{\text{Dirac}}(\epsilon)
&= -s\frac{1}{\pi}\Im\int\frac{d\bm{k}}{(2\pi)^3}(G^s_c+G^s_v)\label{dosdirac},
\end{align}
and
\begin{align}
D_{\text{flat}}(\epsilon)&=-s\frac{1}{\pi}\Im\int\frac{d\bm{k}}{(2\pi)^3}G^s_0.
\label{dosflat}
\end{align}

Also, we decompose the conductivity into the intraband effect and the interband effect. The intraband effect of the Dirac cone are defined by
\begin{align}
\sigma_{\text{intra}}(\epsilon) &= -\frac{\hbar e^2 v^2}{4\pi}\sum _{s,s'=\pm1}ss'\int\frac{d\bm{k'}}{(2\pi)^3}
\bigl(S_{cc}G^s_c J^{ss'}_{cc}G^{s'}_c\nonumber\\&+S_{vv} G^s_v J^{ss'}_{vv} G^{s'}_v  \bigr).
\label{intra}
\end{align}
And the interband effecs between the Dirac cone and the flat band are defined by
\begin{align}
\sigma_{\text{inter}}(\epsilon) &= -\frac{\hbar e^2 v^2}{4\pi}\sum _{s,s'=\pm1}ss'\int\frac{d\bm{k'}}{(2\pi)^3}
\bigl(S_{0c} G^s_c J^{ss'}_{c0} G^{s'}_0\nonumber\\&+S_{c0} G^s_0 J^{ss'}_{0c} G^{s'}_c +S_{0v} G^s_v J^{ss'}_{v0} G^{s'}_0 \nonumber\\&+S_{v0} G^s_0 J^{ss'}_{0v} G^{s'}_v \bigr).
\label{inter}
\end{align}
The intraband effect of the flat band is zero because the flat band has zero group velocity.
The interband effect between the conduction and valence bands is also zero because of skew symmetry of the Hamiltonian represented by $\hat{\mathcal{H}}^* = - \hat{\mathcal{H}}$ \cite{Habe2019-ss}.


\subsection{\label{calculation}Numerical Calculations}

The self-consistent equations Eqs.~(\ref{green function}), (\ref{self energy}), and (\ref{Bethe}) are solved by numerical iteration \cite{Noro2010-cm}.
We discretize the wavenumber as
\begin{align}
dk_j &=  k_{\text{c}}\frac{j}{\sum_{j=1}^{j_{\text{max}}}j}\label{numerical},
\quad
k_j = \frac{1}{2}dk_j + \sum_{j'=1}^{j-1}dk_{j'},
\end{align}
where $j=1,2,...,j_{\text{max}}$ and $k_{\text{c}}$ is the cutoff wavenumber.
Hereafter, we fix $j_{\text{max}}=100$.

\section{\label{linear2} Density of states and conductivity}
The DOS and conductivity are obtained by SCBA. 
Note that the results do not explicitly depend on $q_0$ because the DOS and the conductivity are functions of $\epsilon/(q_0 \hbar v)$ and $\alpha$, and is normalized by $q_0^2/\hbar v$ and $e^2 q_0/\hbar$, respectively.
We fix $k_{\text{c}}=q_0$ in the following.

\subsection{\label{result_scba}Density of states}
\begin{figure*}
\includegraphics[width=16cm]{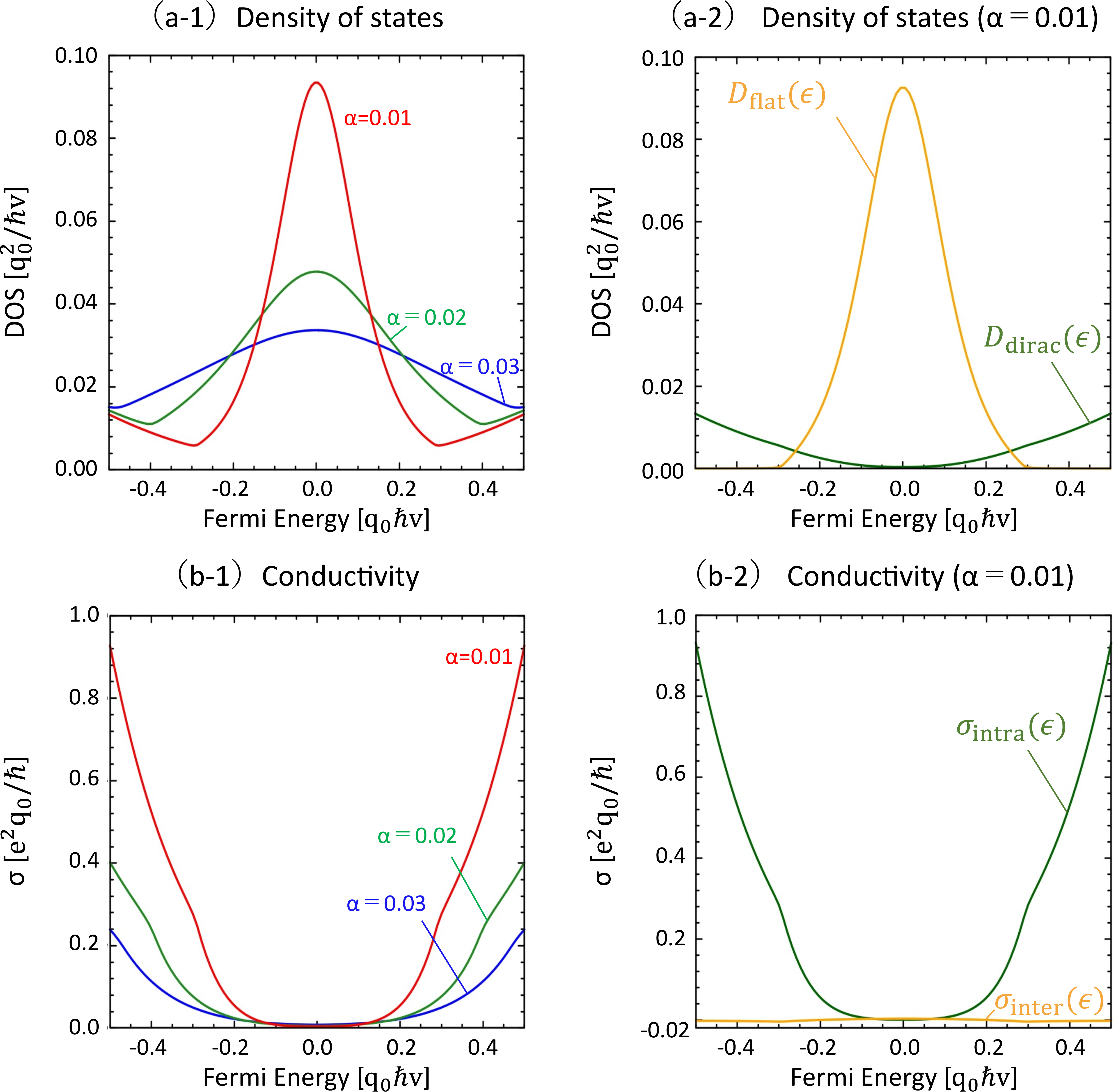}
\caption{(Color online) Quantum transport for $k_{\text{c}}=q_0$ derived by the SCBA. 
	(a-1) DOS for $\alpha=0.01$ (red line), $\alpha=0.02$ (green line), and $\alpha=0.03$ (blue line). 
	(a-2) DOS from the Dirac cone (green line) and the flat band (yellow line) for $\alpha=0.01$.
	(b-1) The conductivity for $\alpha=0.01$ (red line), $\alpha=0.02$ (green line), and $\alpha=0.03$ (blue line). 
	(b-2) The conductivity from the intraband contribution of the Dirac cone (green line) and from the interband contribution between the Dirac cone and the flat band (yellow line) for $\alpha=0.01$.}
\label{results}
\end{figure*}

Figure~\ref{results}(a-1) shows the DOS as a function of Fermi energy.
A peak structure is observed around the zero energy, which becomes broad as $\alpha$ increases.
In the high energy region, it approaches $\epsilon^2/2\pi^2(\hbar v)^3$, the same as the clean limit.
Figure \ref{results}(a-2) provides a detailed understanding of these behaviors; the DOS of the Dirac cones [Eq.~(\ref{dosdirac})] and the flat band [Eq.~(\ref{dosflat})].
The DOS from the flat band makes a pronounced peak with a sharp onset at $\epsilon \sim 0.3$, while the DOS from the Dirac cone is nearly proportional to $\epsilon^2$, similar to that in the clean limit. 
This indicates that the peak in the DOS originates solely from the flat band located at $\epsilon = 0$. 
The Dirac-cone contribution dominates the DOS in the high-energy region.

\subsection{Conductivity}\label{sec:SCBAconductivity}

Figure \ref{results}(b-1) shows the conductivity as a function of Fermi energy.
The conductivity shows a kinked structure at the energy where the onset of the flat-band DOS $D_{\rm flat}$ occurs. 
Namely, the conductivity is slightly suppressed in the low-energy region inside the kinks. 
In this region, the flat-band states, which have the zero group velocity, are dominant hence the conductivity is suppressed. 
Also, as $\alpha$ increases and the peak of the DOS broadens, the positions of the kinks move to a higher energy.

Figure \ref{results}(b-2) shows the results of decomposing the conductivity for $\alpha=0.01$ into the intraband [Eq.~(\ref{intra})] and the interband contributions between the Dirac cone and flat band [Eq.~(\ref{inter})]. 
The contribution of the interband effect is comparable to that of the intraband effect in the low-energy region.
On the one hand, in the high-energy region, the intraband term becomes dominant.

\begin{figure}
\includegraphics[width=7cm]{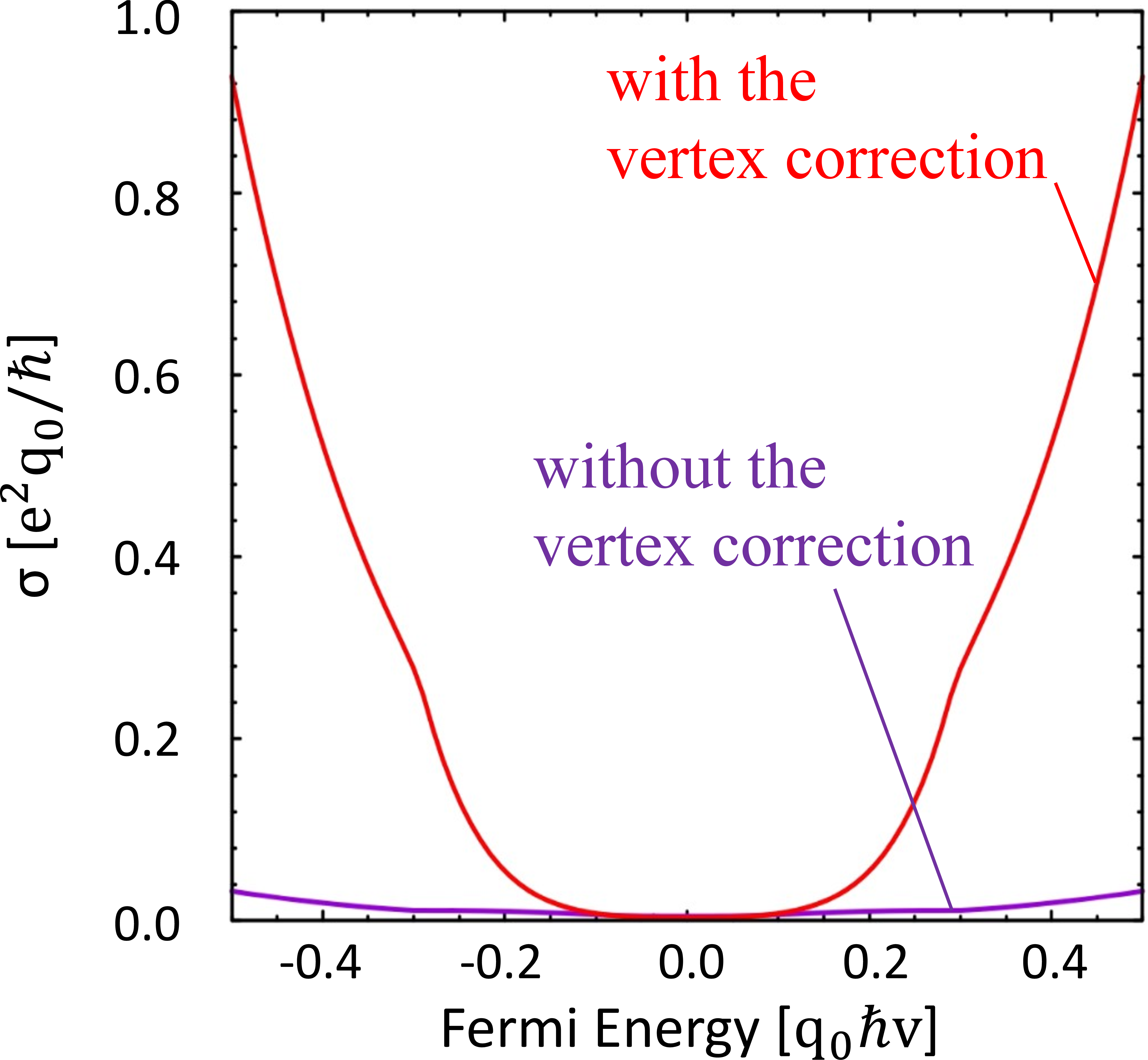}
\caption{(Color online) The conductivity for $\alpha=0.01$ and $k_{\text{c}}=q_0$ derived by the SCBA with the vertex correction (red line) and without the vertex correction (purple line).}
\label{with-vertex} 
\end{figure}

In addition, we find that almost only the vertex correction contributes to the conductivity in the high-energy region, as shown in Fig.~\ref{with-vertex} and will be discussed in detail in Sec.~\ref{scattering_angle}.

\subsection{\label{sec:level2}Screening effect}\label{Cutoff dependence}
\begin{figure*}
\includegraphics[width=16cm]{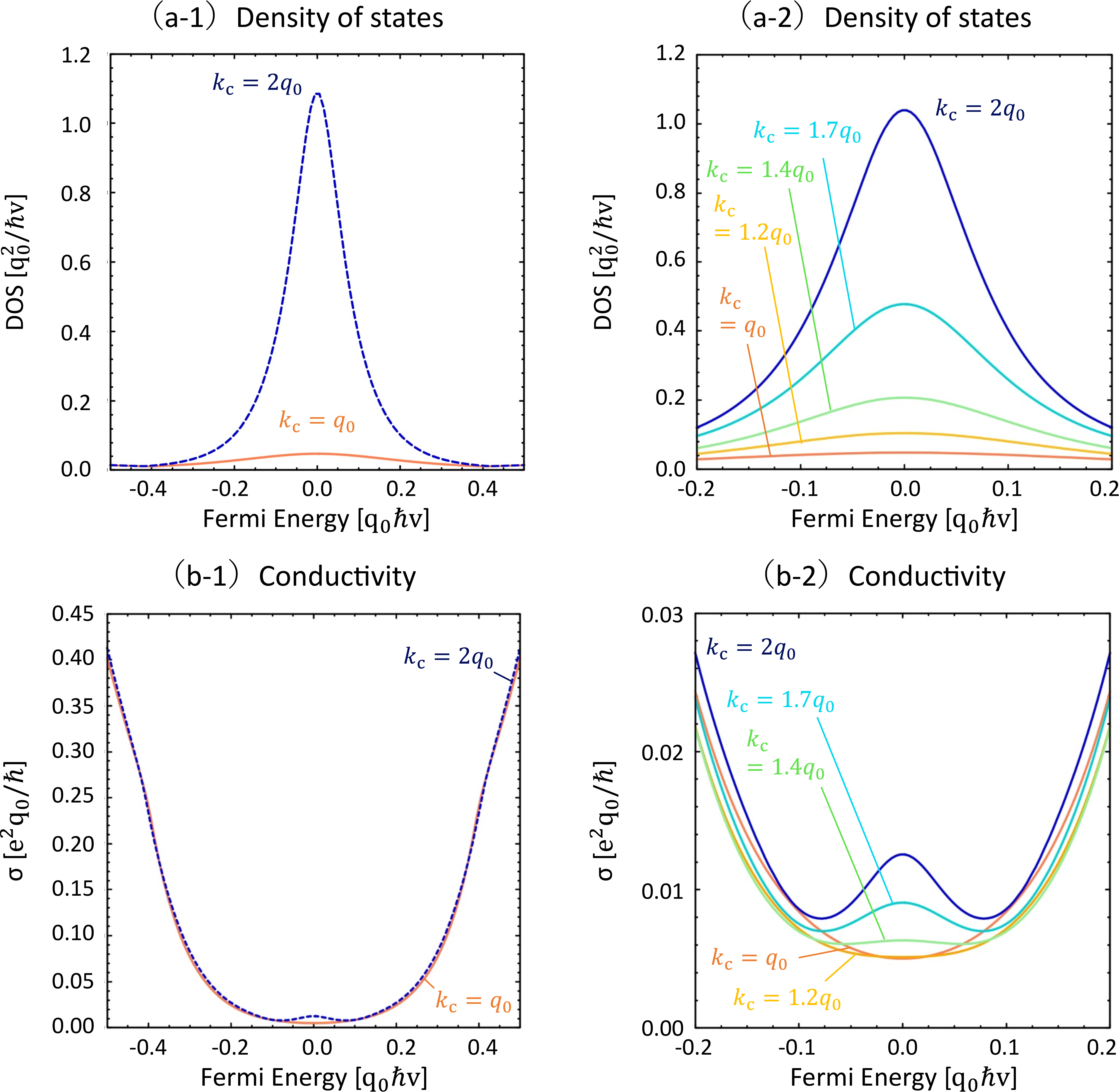}
\caption{(Color online) Quantum transport for $\alpha=0.02$ derived by the SCBA. 
(a-1) DOS and (b-1) conductivity for $k_{\text{c}}=q_0$ (orange solid line) and $k_{\text{c}}=2q_0$ (blue dashed line). 
(a-2) DOS and (b-2) conductivity in the low energy region for $k_{\text{c}}=q_0$ (orange line), $k_{\text{c}}=1.2q_0$ (yellow line), $k_{\text{c}}=1.4q_0$ (green line), $k_{\text{c}}=1.7q_0$ (light-blue line) and $k_{\text{c}}=2q_0$ (blue line).}
\label{kc}
\end{figure*}

\begin{figure*}
\includegraphics[width=17.5cm]{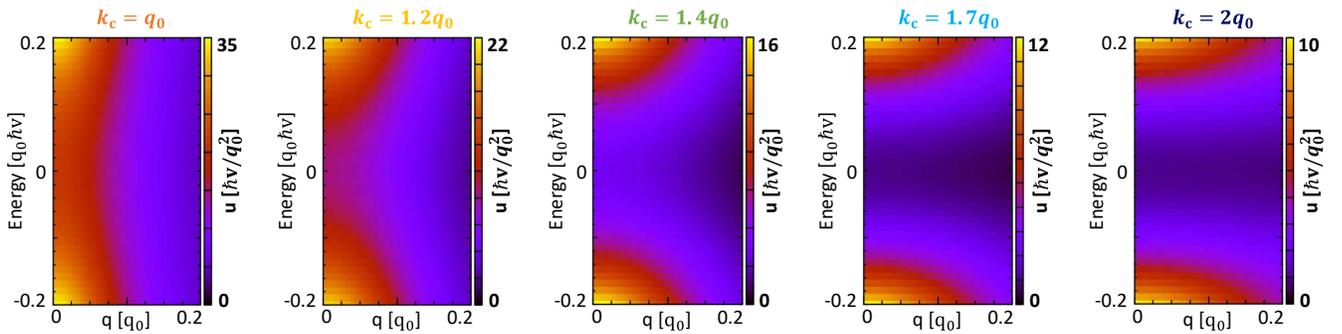}
\centering
\caption{(Color online) The impurity potential $u(\bm{q})$ for $\alpha=0.02$ as a function of $q$ and the Fermi energy derived by the SCBA. 
	From left to right, the cutoff wavenumber is taken as $k_{\text{c}}=q_0$, $k_{\text{c}}=1.2q_0$, $k_{\text{c}}=1.4q_0$, $k_{\text{c}}=1.7q_0$ and $k_{\text{c}}=2q_0$.}
\label{uq}
\end{figure*}

\begin{figure*}
\includegraphics[width=16cm]{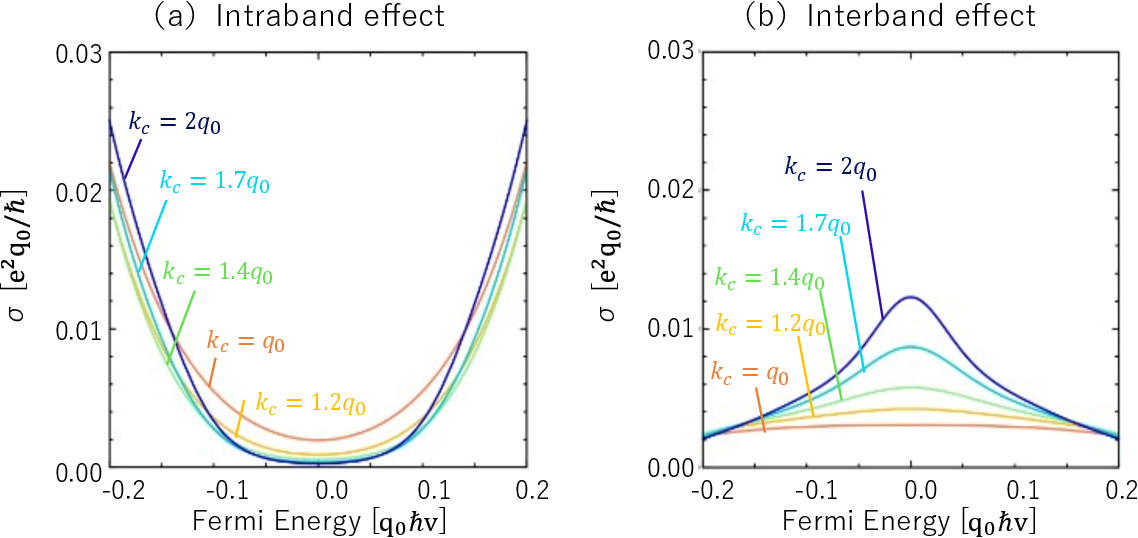}
\centering
\caption{(Color online) The conductivity for $\alpha=0.02$ derived by the SCBA (a) from the intraband contribution of the Dirac cone and (b) from the interband contribution between the Dirac cone and the flat band, for $k_{\text{c}}=q_0$ (orange line), $k_{\text{c}}=1.2q_0$ (yellow line), $k_{\text{c}}=1.4q_0$ (green line), $k_{\text{c}}=1.7q_0$ (light-blue line) and $k_{\text{c}}=2q_0$ (blue line). This is a decomposition of Fig.~\ref{kc}(b-2) into intraband and interband effects.}
\label{band}
\end{figure*}

The significance of the screening effect is depicted in Figs.~\ref{kc} and \ref{uq}, where the dependence of the DOS and conductivity on the cutoff wavenumber $k_{\rm c}$ is demonstrated. 
More details are provided below. 
For these analyses, we have set $\alpha=0.02$.

Figures~\ref{kc}(a-1) and \ref{kc}(b-1) show the DOS and conductivity for the cutoff wavenumber $k_{\text{c}}=q_0$ and $2q_0$.
The larger $k_{\text{c}}$ is, the higher the peak of the DOS becomes because the number of flat-band states considered increases.
On the other hand, the conductivity shows a small peak near zero Fermi energy for $k_{\text{c}}=2q_0$. 
The evolution of DOS and conductivity for $-0.2<\epsilon<0.2$ with increasing $k_{\rm c}$ is shown in Figs.~\ref{kc}(a-2) and \ref{kc}(b-2).
It can be seen that the conductivity near zero Fermi energy increases as $k_{\text{c}}$ increases. 

As shown in Fig.~\ref{uq}, the Coulomb impurity potential has a strong Fermi energy dependence. The potential is suppressed in the low-energy region for a large $k_{\rm c}$.
The large DOS for a large $k_{\rm c}$ in the low-energy region as in Fig.~\ref{kc}(a-2) enhances the screening effect, resulting in a large value of $q_{\text{s}}$ within the Thomas-Fermi approximation Eq.~(\ref{Thomas-Fermi}). 
Then, the larger the value of $k_{\text{c}}$, the smaller the impurity potential becomes in the low-energy region.
In the high-energy region, the DOS is not substantially enhanced. 
Thus, the conductivity generates a peak when $k_{\text{c}}$ is significant, even though flat-band states have a vanishing group velocity in the vicinity of zero Fermi energy. 
As shown in Fig~\ref{band}, decomposition of conductivity shown in Fig.~\ref{kc}(b-2) into intraband [Fig.~\ref{band}(a)] and interband  [Fig.~\ref{band}(b)] effects, the interband term results in the formation of the peak around zero energy. On the contrary, the intraband term is nearly independent of $k_{\rm c}$.

\section{\label{Boltzmann}Boltzmann transport theory}
The Boltzmann equation can be used to derive the qualitative behavior of the intraband effect more easily than SCBA. 
Here, we derive the conductivity from the Boltzmann equation to compare the results of the SCBA and the Boltzmann equation and to explain why the vertex correction effect is more significant.

\subsection{Conductivity}

The conductivity at the zero temperature is given by
\begin{align}
	\sigma_{\mathrm{B}}(\epsilon)&= \frac{e^2v^2}{3}D_0(\epsilon)\tau_{\mathrm{tr}}(\epsilon),
	\label{conductivity_B}
\end{align}
with $D_0$ the DOS per unit volume in the clean limit and $\tau_{\mathrm{tr}}$ the transport relaxation time.
The DOS is given by 
\begin{align}
	D_0(\epsilon)=\frac{\epsilon^2}{2\pi^2(\hbar v)^3}
	\qfor \epsilon \neq 0,
	\label{dos0}
\end{align}
originating from the Dirac cone ($\lambda=\pm1$).
The transport relaxation time is calculated by
\begin{align}
	\frac{1}{\tau_{\rm{tr}}(\epsilon_{\lambda,\bm{k}})} = \sum_{\lambda'}\int\frac{d\bm{k'}}{(2\pi)^3}(1-\cos\theta_{\bm{k'}\bm{k}})W_{\lambda'\bm{k'}, \lambda\bm{k}}. \label{tau}
\end{align}
The scattering probability $W_{\lambda'\bm{k}',\lambda \bm{k}}$ is given by the Fermi's golden rule as
\begin{align}
W_{\lambda'\bm{k'}, \lambda\bm{k}} = \frac{2\pi}{\hbar}n_{\text{i}}|
\bra{\lambda',\bm{k'}}U\ket{\lambda,\bm{k}}|^2\delta(\epsilon_{\lambda',\bm{k'}}-\epsilon_{\lambda,\bm{k}}),\label{golden_rule}
\end{align}
where
\begin{align}
 \bra{\lambda',\bm{k'}}U\ket{\lambda,\bm{k}}
 = \int d\vb*{r} e^{-i (\vb*{k}-\vb*{k}') \cdot \vb*{r}}
 U(\vb*{r}) \vb*{v}_{\lambda', \vb*{k}}^\dag \vb*{v}_{\lambda, \vb*{k}}.
\end{align}
The transport relaxation time is obtained as
\begin{align}
\frac{1}{\tau_{\mathrm{tr}}(\epsilon)}
&=\frac{n_{\text{i}}}{2\pi\hbar^2 v}\int_{-1}^1\int_{0}^{\infty} dk'd(\cos\theta_{\bm{kk'}})k'^2(1-\cos\theta_{\bm{kk'}})
\nonumber\\&\quad\times
\frac{(\cos\theta_{\boldsymbol{kk}'}+1)^2}{4}\delta(k-k')|u(\bm{k}-\bm{k'})|^2
\\ & \hspace{-1em}
=\frac{n_{\text{i}}\epsilon^2}{4(2\pi)^2\hbar(\hbar v)^3}
\bigl[V_0^2(\epsilon/\hbar v,\epsilon/\hbar v)+V_1^2(\epsilon/\hbar v,\epsilon/\hbar v)
\notag\\&
-V_2^2(\epsilon/\hbar v,\epsilon/\hbar v)-V_3^2(\epsilon/\hbar v,\epsilon/\hbar v)
\bigr],
\label{conductivity_B_spin1}
\end{align}
where the momentum is set to the Fermi momentum, $k=\epsilon/\hbar v$. 
\begin{figure}
\includegraphics[width=7cm]{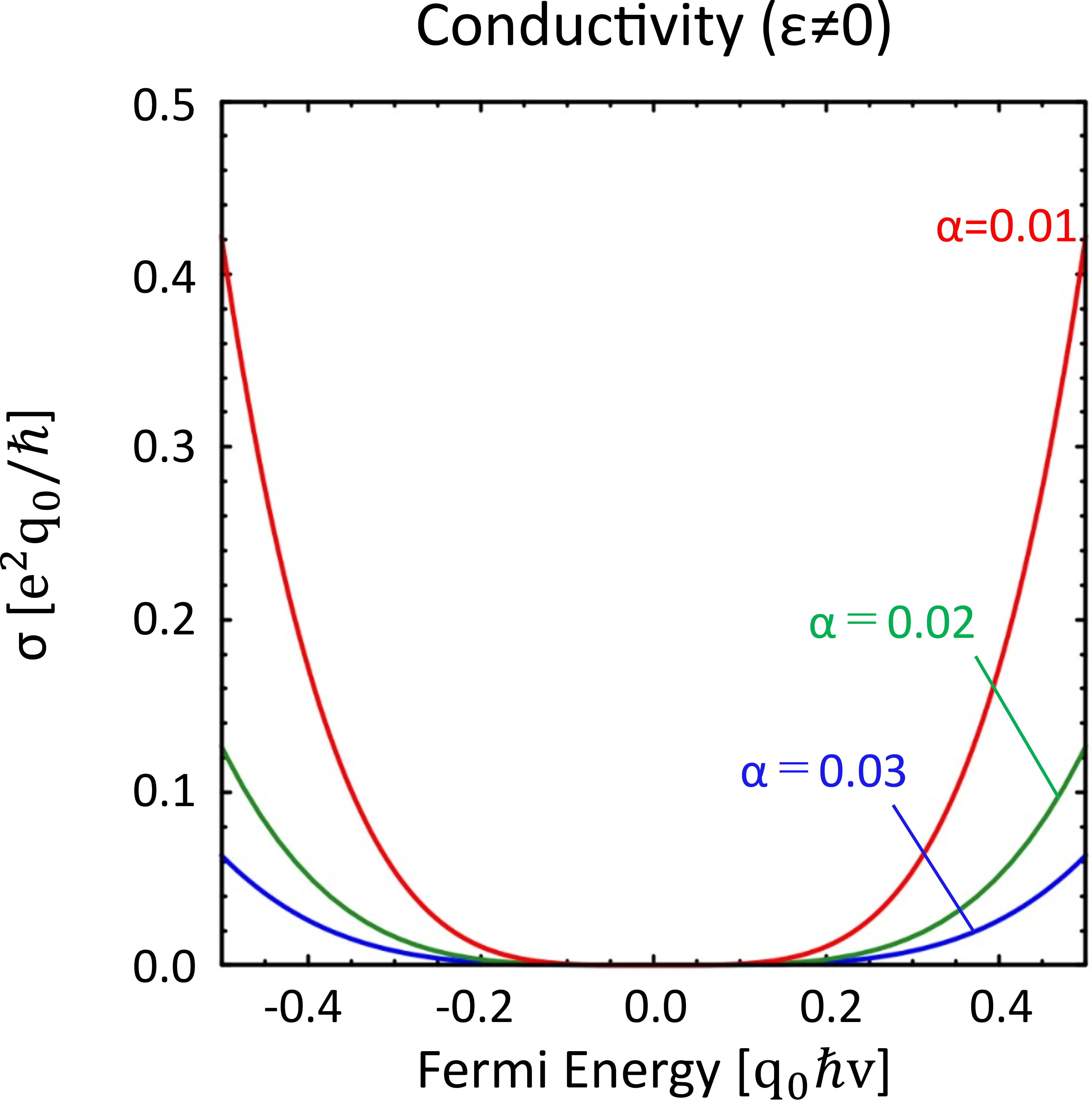}
\caption{(Color online) Electrical conductivity of the spin-1 fermion for $\alpha=0.01$ (red line), $\alpha=0.02$ (green line), and $\alpha=0.03$ (blue line), derived by the Boltzmann equation for $\epsilon \neq 0$.}
\label{bol}
\end{figure}
As a result, we find the conductivity 
\begin{align}
&\sigma_\text{B}(\epsilon)=\frac{e^2\epsilon^4}{6\pi\hbar^5v^4q_0^3\alpha^2}\times
\notag\\
&\frac{1}{-\pi(5\pi+3\alpha)+(2\pi+\alpha)(2\pi+3\alpha)\text{arctanh} \left(\displaystyle\frac{\pi}{\pi+\alpha}\right)},
\label{C-Bol-c}
\end{align}
for the Coulomb potential. 
This result is shown in Fig.~\ref{bol}. 

\subsection{Comparison of results from Boltzmann equation and SCBA}
\begin{figure}
\includegraphics[width=7cm]{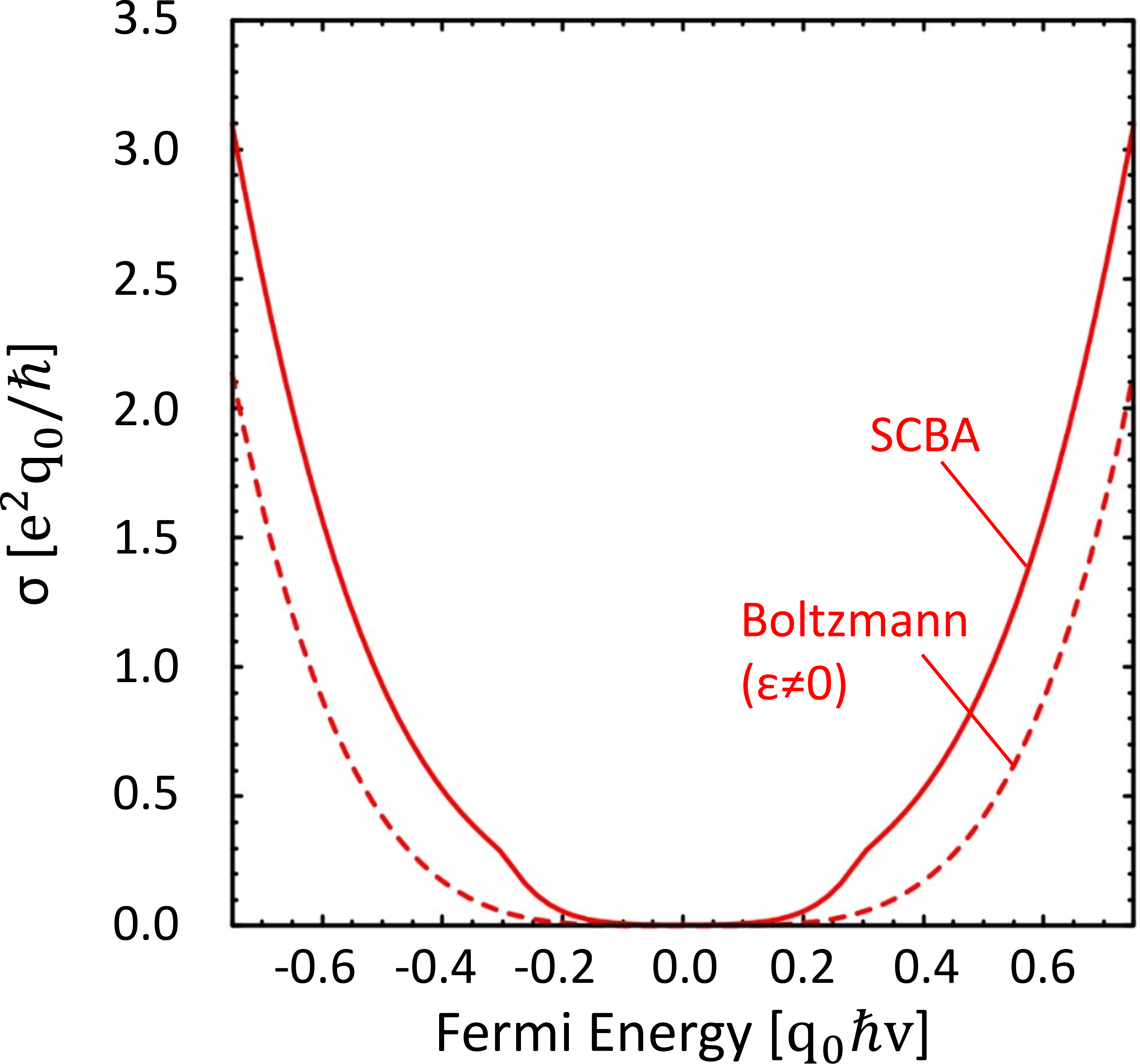}
\caption{(Color online) The conductivity ($\alpha=0.01$) derived by the Boltzmann equation (dashed line) for $\epsilon \neq 0$ and the SCBA (solid line) for $k_{\text{c}}=q_0$.}
\label{compare} 
\end{figure}

Figure~\ref{compare} compares the conductivity derived from the Boltzmann equation with that derived from SCBA.
As the Boltzmann equation does not consider the significant contribution of the flat band with spectral broadening, it cannot reproduce the kink-like structure seen in the SCBA for $\epsilon \sim 0.3$, a point we elaborated in Sec.~\ref{sec:SCBAconductivity}.

\subsection{Scattering angle dependence}
\label{scattering_angle}

The conductivity obtained by SCBA shows that the vertex correction effect contributes significantly to the conductivity, as described in Sec.~\ref{sec:SCBAconductivity}.
The vertex correction takes into account the scattering angle dependence of the conductivity. 
That is, the vertex correction incorporates the effect that forward scattering has a small contribution to conductivity suppression and backward scattering has a large contribution.
As a result, the conductivity experiences a substantial increase due to the vertex correction, particularly when forward scattering is the dominant process. 

\begin{figure}
\includegraphics[width=7cm]{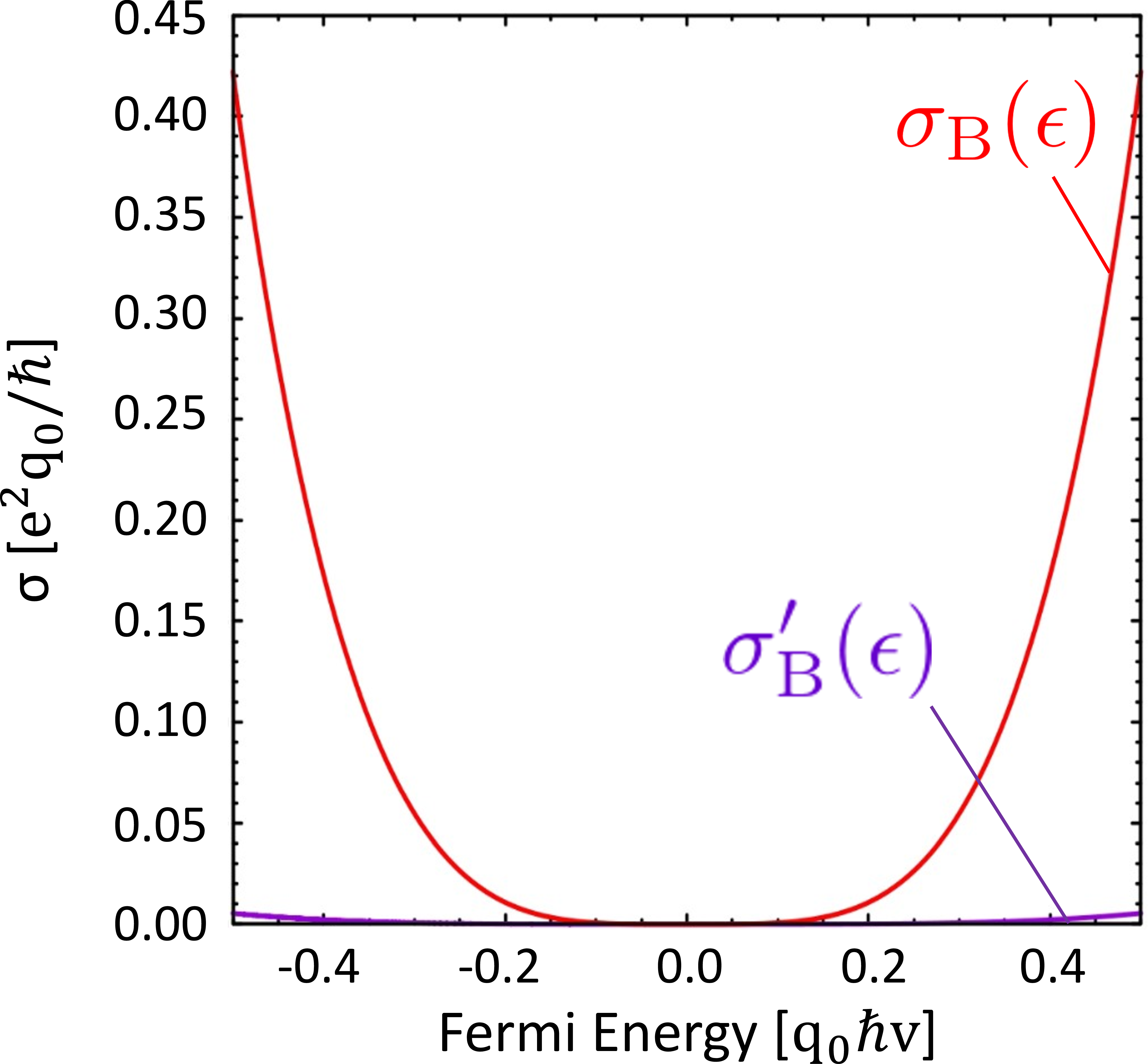}
\caption{(Color online) The conductivity ($\alpha=0.01, \epsilon\neq 0$) derived by the Boltzmann equation. $\sigma_\text{B}(\epsilon)$ (red line) derived including the second term in Eq.~(\ref{tau}), and $\sigma'_\text{B}(\epsilon)$ (purple line) derived not including it.}
\label{bol_ver} 
\end{figure}

In the case of Boltzmann equation, on the other hand, the second term in Eq.~(\ref{tau}) corresponds to the effect of the scattering angle dependence. 
%
To extract the contribution from the scattering angle dependence, we define the momentum relaxation time $\tau'_{\mathrm{tr}}(\epsilon)$, not including the second term in Eq.~(\ref{tau}), is
\begin{align}
\frac{1}{\tau'_{\mathrm{tr}}(\epsilon)}
&=\frac{n_{\text{i}}}{2\pi\hbar^2 v}\int_{-1}^1\int_{0}^{\infty} dk'd(\cos\theta_{\bm{kk'}})k'^2
\nonumber\\&\quad\times
\frac{(\cos\theta_{\boldsymbol{kk}'}+1)^2}{4}\delta(k-k')|u(\bm{k}-\bm{k'})|^2
\notag\\ & \hspace{-1em}
=\frac{n_{\text{i}}\epsilon^2}{4(2\pi)^2\hbar(\hbar v)^3}
\bigl[V_0^2(\epsilon/\hbar v,\epsilon/\hbar v)+2V_1^2(\epsilon/\hbar v,\epsilon/\hbar v)
\notag\\&
+V_2^2(\epsilon/\hbar v,\epsilon/\hbar v)
\bigr].
\end{align}
The corresponding conductivity $\sigma'_\text{B}(\epsilon) = e^2 v^2 D_0(\epsilon) \tau_{\mathrm{tr}}'(\epsilon)/3$ is obtained as
\begin{align}
&\sigma'_\text{B}(\epsilon)
=
\nonumber\\
& \frac{e^2\epsilon^4}{12\pi^2\hbar^5v^4q_0^3\alpha
	\left[\pi(\pi+\alpha)-\alpha(2\pi+\alpha)\text{arctanh} \left(\displaystyle\frac{\pi}{\pi+\alpha}\right)\right]}.
\end{align}
The comparison between $\sigma_{\mathrm{B}}$ and $\sigma_{\mathrm{B}}'$ is shown in Fig.~\ref{bol_ver}. 
This result suggests that the forward scattering dominates over the scattering process under the Coulomb-potential impurity, implying the vertex correction is substantial.

\begin{figure}
\includegraphics[width=7cm]{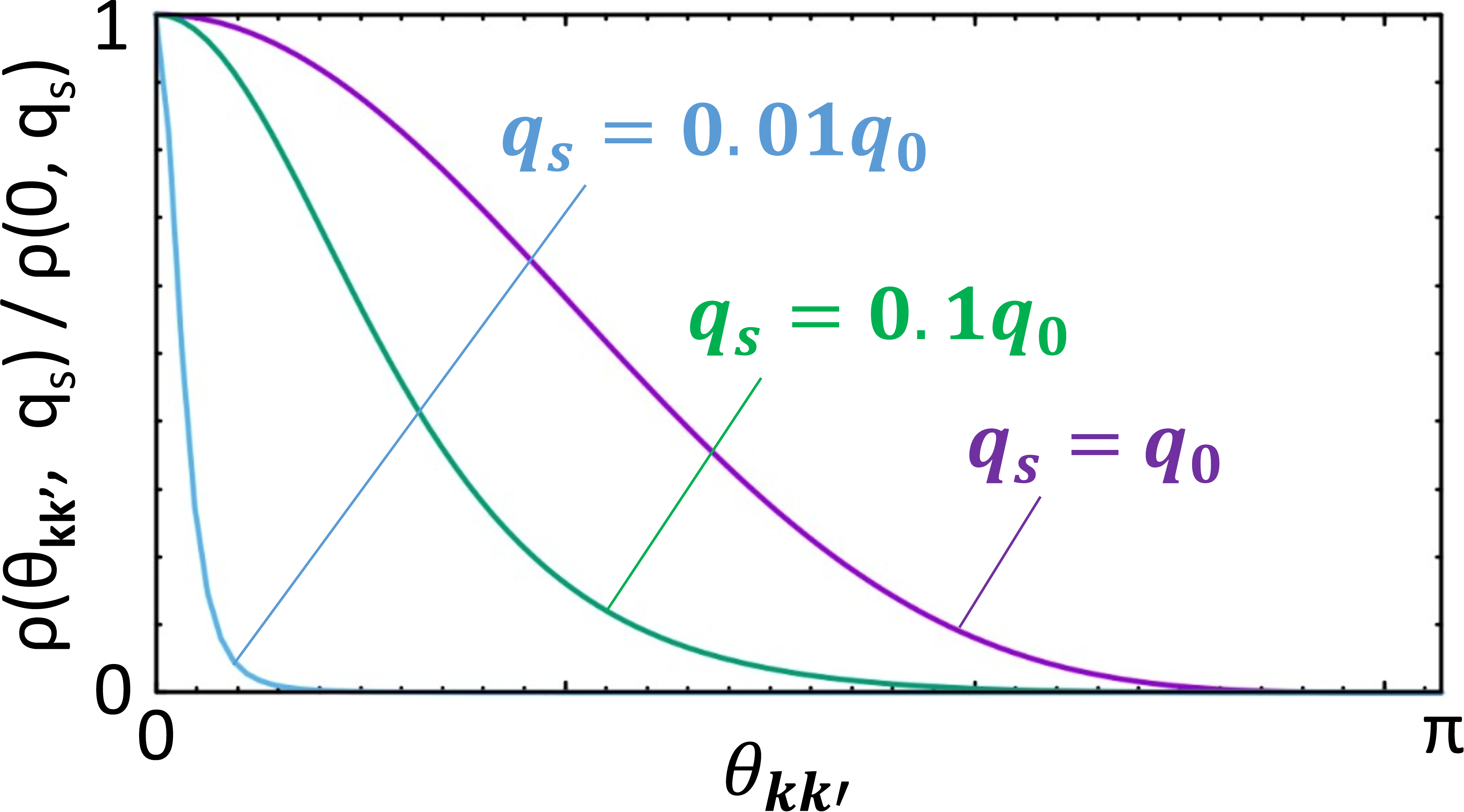}
\caption{(Color online) 
	The {$\theta_{\bm{kk'}}$} dependence of the scattering amplitude $\rho(\theta_{\bm{kk'}}, q_{\text{s}})$ for $q_{\text{s}}=q_0$ (purple), $q_{\text{s}}=0.1q_0$ (green), $q_{\text{s}}=0.01q_0$ (light blue).}
\label{fig_qs} 
\end{figure}

The discussions presented above are corroborated by investigating the angle-dependent scattering amplitude associated with the Coulomb potential.
For a spin-1 fermion, the scattering amplitude $\rho(\theta_{\bm{k}\bm{k'}}, q_{\text{s}})$ shows an anomalous anisotropy
\begin{align}
\rho(\theta_{\bm{k}\bm{k'}}, q_{\text{s}}) 
&= |\bra{\pm 1,\bm{k}}U\ket{\pm 1,\bm{k}}|^2\label{rho}
\notag\\
&= |u(\bm{k})|^2|\bra{\pm 1,\bm{k}}\ket{\pm 1,\bm{k}}|^2
\notag\\
&= \left(\frac{4\pi e^2}
{\kappa\left(4k^2\sin^2 {\theta_{\bm{k}\bm{k'}}}/{2} + q_{\text{s}}^2\right)}\right)^2
\cos^4\frac{\theta_{\bm{k}\bm{k'}}}{2},
\end{align}
as depicted in Fig.~\ref{fig_qs}. 
As $q_{\text{s}}$, the inverse of the Thomas-Fermi screening length, decreases, the forward scattering $\theta_{\bm{k}\bm{k'}} \sim 0$ contribution increases and diverges as $\sin^{-4}\theta_{\bm{kk'}}/2$ in the unscreened limit $q_{\mathrm{s}} \to 0$. 
For a typical value, $q_{\text{s}}\sim 0.01q_0$ for $\alpha=0.01$, the contribution predominantly stems from forward scattering.

Thus, in the present system, the vertex correction in the SCBA and the second term in Eq.~(\ref{tau}) within the Boltzmann equation significantly contribute to the conductivity. 
This is due to the small screening effect and the dominant forward scattering.
This fact for the spin-1 fermion is obtained by incorporating the screening effect from the DOS in a self-consistent manner. 
A similar significant vertex correction effect can be anticipated for topological semimetals with a small Fermi energy.

\section{\label{discussion}Discussion}
We studied the transport phenomena of spin-1 fermions under Coulomb-type impurity potential and observed a peak in the DOS and a suppression of conductivity, similar to the results previously obtained for Gaussian impurity potential \cite{Kikuchi2022}. 
On the other hand, when comparing the conductivity under the two types of potentials, we found a noticeable difference in the dependence from the number of flat-band states (cutoff wavenumber), as discussed below. 
We also comment on the experimental realization of our result.

\subsection{Conductivity depending on the number of flat-band states (cutoff wavenumber)}

The conductivity for the Coulomb potential increases in the low-energy region as the number of flat-band states, proportional to the cutoff wavenumber $k_{\rm c}$, increases. 
In contrast, when the screening effect is not considered self-consistently, the conductivity for the Gaussian potential in the low-energy region approaches converges in the $k_{\rm c} \to \infty$ limit (see Appendix \ref{appendix:gauss}).
The self-consistent formalism allows for the consideration of the amplified screening effect in the low-energy region. 
This enhancement is facilitated by the disorder potential increasing the DOS, which in turn results in an increase in conductivity.
Therefore, the results differ from those calculated for the Gauss potential without the self-consistent Fermi-energy dependence of the screening effect.

In typical metals, on the one hand, the Fermi-energy dependence on DOS is negligibly small the conductivity is insensitive to the self-consistent inclusion of the screening effect.

\subsection{Thomas-Fermi approximation}
In the present model of the Coulomb potential, the Thomas-Fermi approximation is considered by taking the long-wavelength limit $q \to 0$ of the polarization function. 
Note that in the $q \to 0$ limit Eqs.~(\ref{Thomas-Fermi}) and (\ref{u_the Coulomb}) are applicable to any system due to the compressibility sum rule. 
On the other hand, it has been shown that the polarization function of spin-1 fermions increases with $q$ \cite{Dey2022-th} and implies that the screening effect may be underestimated by the Thomas-Fermi approximation \cite{Noro2010-cm}.
Considering the full polarization function could yield better quantitative results, which is an issue for future investigation.

\subsection{On experiment}
We theoretically clarified the dependence of the DOS and the conductivity on the Fermi energy. 
These are measured experimentally by continuously varying the Fermi energy with the gating in the thin films of the spin-1 fermion materials. 
Alternatively, the Fermi energy can be varied discretely by doping the bulk material. 
%
However, besides spin-1 fermions, actual materials also possess electronic states characterized by double Weyl fermions. To fully comprehend transport measurements, it is necessary to understand the quantum transport phenomena associated with double Weyl fermions. 

It is useful to evaluate the parameters and the order of conductivity for further theoretical and experimental studies. 
To date, no estimation has been made for the value of $\kappa$ in spin-1 fermion systems. 
However, the Dirac semimetal $\text{Cd}_3\text{As}_2$ has $\kappa = 36$ \cite{J.-P.Jay-Gerin1977}. 
Thus, we assume that a topological semimetal with a spin-1 fermion could exhibit $\kappa = 10$, comparable to that of $\text{Cd}_3\text{As}_2$.
The effective fine-structure constant $\alpha$ is $0.01\sim 0.1$  for $\kappa \sim 10$.
The cutoff wavenumber $k_\text{c}$ is of the order of the reciprocal of the lattice constant, $k_\text{c}\sim 1 \, \text{\AA}^{-1}$. 
When the impurity concentration is about $0.1\%$, the characteristic wavenumber is $q_0 = n_{\rm i}^{1/3} \sim 0.1 \, \text{\AA}^{-1}$. 
The unit of energy is approximately $q_0 \hbar v \sim 1 \, \text{eV}$, the unit of DOS is $q_0^2/\hbar v\sim 10^{-3} \, \text{eV}^{-1} \, \text{\AA}^{-3}$, and the unit of conductivity is $e^2 q_0/\hbar \sim 1 \, \text{m}\Omega^{-1} \, \text{cm}^{-1}$, which is of the order of the conductivity in semimetals.

%

\section{\label{conclusion}Conclusion}
This study has elucidated the quantum transport theory for a spin-1 chiral fermion under the Coulomb impurities within the Self-Consistent Born Approximation (SCBA), accounting for the current vertex correction. Consequently, we observed a peak structure in the density of states and identified the suppression of conductivity arising from the flat band near zero energy. Additionally, we discovered that an increase in the number of flat band electrons (resulting from an increased cutoff wavenumber) leads to an anomalously strong screening effect and substantial electrical conductivity in the low-energy region. 
This is attributed to the pronounced Fermi-energy dependence of the screening length for the Coulomb potential. 
In contrast, no such effect is observed for the Gaussian impurity potential without the screening effect derived by a self-consistent method. Furthermore, our findings indicate that most of the scattering is forward scattering, which amplifies the vertex correction effect.
The results of this study provide a foundation for understanding the quantum transport behavior of spin-1 fermions. 
The research suggests the potential existence of nontrivial impurity effects in multifold fermions. Exploring the quantum transport phenomena in various chiral-fermion systems presents an intriguing area for future research.

\begin{acknowledgments}
	This work is supported by JSPS KAKENHI for Grants (Grants No. JP20K03835) and the Sumitomo Foundation (Grant No. 190228).
    RK would like to take this opportunity to thank the “Nagoya University Interdisciplinary Frontier Fellowship” supported by Nagoya University and JST, the establishment of university fellowships towards the creation of science technology innovation, Grant Number JPMJFS2120.
\end{acknowledgments}

\appendix

\section{Detailed calculations}\label{j}

The self-energy is expressed as
\begin{align}
\hat{\Sigma}(\bm{k},\epsilon)
 &= \Sigma_1(k,\epsilon)\hat{S}_0
 +\Sigma_2(k,\epsilon)(\hat{\bm{S}}\cdot\bm{n})+\Sigma_3(k,\epsilon)(\hat{\bm{S}}\cdot\bm{n})^2,
 \label{self energy2}
\end{align}
because of $(\hat{\bm{S}}\cdot\bm{n})^3=(\hat{\bm{S}}\cdot\bm{n})$.
From the above expression, Eq.~(\ref{green function}) can be rewritten as
\begin{align}
\hat{G}(\bm{k},\epsilon)
&= \frac{1}{X(k,\epsilon) \hat{S}_0+Y(k,\epsilon) \hat{\bm{S}}\cdot\bm{n}+Z(k,\epsilon) (\hat{\bm{S}}\cdot\bm{n})^2}
\notag\\
&= x(k,\epsilon) \hat{S}_0+y(k,\epsilon) (\hat{\bm{S}}\cdot\bm{n})+z(k,\epsilon) (\hat{\bm{S}}\cdot\bm{n})^2\label{green function2},
\end{align}
where
\begin{align}
X(k,\epsilon)&=\epsilon-\Sigma_1(k,\epsilon),
\\
Y(k,\epsilon)&=-\hbar vk-\Sigma_2(k,\epsilon),
\\
Z(k,\epsilon)&=-\Sigma_3(k,\epsilon),
\end{align}
and
\begin{align}
x(k,\epsilon)&=\frac{1}{X(k,\epsilon)},
\\
y(k,\epsilon)&=-\frac{Y(k,\epsilon)}{(X(k,\epsilon)+Z(k,\epsilon))^2-Y(k,\epsilon)^2},
\\
z(k,\epsilon)&=\frac{Y(k,\epsilon)^2-Z(k,\epsilon)(X(k,\epsilon)+Z(k,\epsilon))}{((X(k,\epsilon)+Z(k,\epsilon))^2-Y(k,\epsilon)^2)X(k,\epsilon)}.
\end{align}
Substituting Eq.~(\ref{green function2}) into Eq.~(\ref{self energy}), we obtain 
\begin{align}
	&
\hat{\Sigma}(\bm{k},\epsilon+is0)
\notag\\ &
= \hat{S}_0\int\frac{k'^2dk'}{(2\pi)^3}n_{\text{i}}\left[V_0^2(k,k')x(k',\epsilon+is0)\right.\nonumber\\
& \hspace{7em}
\left.+(V_0^2(k,k')-V_2^2(k,k'))z(k',\epsilon+is0)\right]\nonumber\\
&\quad
+(\hat{\bm{S}}\cdot\bm{n})\int\frac{k'^2dk'}{(2\pi)^3}n_{\text{i}}V_1^2(k,k')y(k',\epsilon+is0)
\nonumber\\
&\quad
+(\hat{\bm{S}}\cdot\bm{n})^2\int\frac{k'^2dk'}{(2\pi)^3}n_{\text{i}}\left(\frac{3}{2}V_2^2(k,k')-\frac{1}{2}V_0^2(k,k')\right)\nonumber\\
&\hspace{10em}\times 
z(k',\epsilon+is0)\label{self energy3},
\end{align}
using the relations shown in Appendix \ref{integral}.
Comparing Eq.~(\ref{self energy3}) with Eq.~(\ref{self energy2}), we get
\begin{align}
	&
\Sigma_1(k,\epsilon+is0)
 =\int\frac{k'^2dk'}{(2\pi)^3}n_{\text{i}}
 \bigl[V_0^2(k,k')x(k',\epsilon+is0)
\nonumber\\& \hspace{4em} +
(V_0^2(k,k')-V_2^2(k,k'))z(k',\epsilon+is0) \bigr],
\label{self1}
\\ &
\Sigma_2(k,\epsilon+is0)
=\int\frac{k'^2dk'}{(2\pi)^3}n_{\text{i}}V_1^2(k,k')y(k',\epsilon+is0),
\label{self2}
\\&
\Sigma_3(k,\epsilon+is0) = \int\frac{k'^2dk'}{(2\pi)^3}n_{\text{i}}\left(\frac{3}{2}V_2^2(k,k')-\frac{1}{2}V_0^2(k,k')\right)\nonumber\\&
\hspace{12em}
\times z(k',\epsilon+is0).
\label{self3}
\end{align}

Substituting Eq.~(\ref{green function2}) into Eq.~(\ref{dos}), the density of states can be rewritten as
\begin{align}
&D(\epsilon)
= -\frac{1}{\pi}\Im\int\frac{d\bm{k}}{(2\pi)^3}\left(\frac{1}{X(k,\epsilon+i0)} \right.
\nonumber\\&\quad
+\frac{1}{X(k,\epsilon+i0)+Y(k,\epsilon+i0)+Z(k,\epsilon+i0)}\nonumber\\
&\quad\left.
+\frac{1}{X(k,\epsilon+i0)-Y(k,\epsilon+i0)+Z(k,\epsilon+i0)}\right).
\end{align}

In addition, the Bethe-Salpeter equation [Eq.~(\ref{Bethe})] can be reduced to a more manageable form. 
The current vertex $\hat{J}_x(\bm{k},\epsilon,\epsilon')$ is decomposed into eight terms as
\begin{align}
\hat{J}_x(\bm{k},\epsilon,\epsilon')
&=\hat{S}_xJ_0(k,\epsilon,\epsilon')+n_x(\hat{\bm{S}}\cdot\bm{n})^2J_1(k,\epsilon,\epsilon')\nonumber\\
&+n_x(\hat{\bm{S}}\cdot\bm{n})J_2(k,\epsilon,\epsilon')+(\hat{\bm{S}}\cdot\bm{n})^2\hat{S}_xJ_3(k,\epsilon,\epsilon')\nonumber\\
&+\hat{S}_x(\hat{\bm{S}}\cdot\bm{n})^2J_4(k,\epsilon,\epsilon')+(\hat{\bm{S}}\cdot\bm{n})\hat{S}_xJ_5(k,\epsilon,\epsilon')\nonumber\\
&+\hat{S}_x(\hat{\bm{S}}\cdot\bm{n})J_6(k,\epsilon,\epsilon')+n_x\hat{S}_0J_7(k,\epsilon,\epsilon')\label{J},
\end{align}
by using Eqs.~(\ref{int1})--(\ref{int2}) for Eq.~(\ref{Bethe})~(see Appendix~\ref{integral}).

 From Appendix~\ref{integral}, the Bethe-Salpeter equation (Eq.~(\ref{Bethe})) can be rewritten as  
\begin{widetext}
\begin{align}
\left( \begin{array}{c}  J_0\\
 J_1\\
 J_2\\
 J_3\\
 J_4\\
 J_5\\
 J_6\\
 J_7 \end{array}\right)
&=&
\left( \begin{array}{c}  1\\
0\\
0\\
0\\
0\\
0\\
0\\ 
0 \end{array}\right)
+
\int \displaystyle\frac{k'^2dk'}{(2\pi)^3}n_{\text{i}}\begin{pmatrix}
V_0^2 & 0 & \frac{1}{2}V_0^2-\frac{1}{2}V_2^2 & V_0^2-V_2^2 & V_0^2-V_2^2 & 0 & 0 & 0 \\
0 & \frac{5}{2}V_3^2-\frac{3}{2}V_1^2 & 0 & 0 & 0 & 0 & 0 & 0 \\
0 & 0 & \frac{3}{2}V_2^2-\frac{1}{2}V_0^2 & 0 & 0 & 0 & 0 & 0 \\
0 & 0 & 0 & \frac{3}{2}V_2^2-\frac{1}{2}V_0^2 & 0 & 0 & 0 & 0 \\
0 & 0 & 0 & 0 & \frac{3}{2}V_2^2-\frac{1}{2}V_0^2 & 0 & 0 & 0 \\
0 & \frac{1}{2}V_1^2-\frac{1}{2}V_3^2 & 0 & 0 & 0 & V_1^2 & 0 & 0 \\
0 & \frac{1}{2}V_1^2-\frac{1}{2}V_3^2 & 0 & 0 & 0 & 0 & V_1^2 & 0 \\
0 & V_1^2-V_3^2 & 0 & 0 & 0 & 0 & 0 & V_1^2 \\
\end{pmatrix}\hat{T}
\left( \begin{array}{c}  J_0'\\
 J_1'\\
 J_2'\\
 J_3'\\
 J_4'\\
 J_5'\\
 J_6'\\
 J_7'\end{array}\right)\label{vertex},\nonumber\\
\end{align}
\end{widetext}
where $J_i=J_i(k,\epsilon+is0,\epsilon+is'0),J'_i=J_i(k',\epsilon+is0,\epsilon+is'0), V_i^2=V_i^2(k,k'),x=x(k',\epsilon+is0),x'=x(k',\epsilon+is'0)$ and so on, and the matrix $\hat T$ is defined as
\begin{align}
\hat{T}=\begin{pmatrix}
xx' & 0 & 0 & 0 & 0 & 0 & 0 & 0 \\
T_{01} & T_{11} & T_{21} & T_{31} & T_{41} & T_{51} & T_{61} & T_{71} \\
T_{02} & T_{12} & T_{22} & T_{32} & T_{42} & T_{52} & T_{62} & T_{72} \\
zx' & 0 & 0 & T_{33} & 0 & yx' & 0 & 0 \\
xz' & 0 & 0 & 0 & T_{44} & 0 & xy' & 0 \\
yx' & 0 & 0 & yx' & 0 & T_{55} & 0 & 0 \\
xy' & 0 & 0 & 0 & xy' & 0 & T_{66} & 0 \\
0 & 0  & 0  & 0  & 0  & 0  & 0  & xx' \\
\end{pmatrix}.
\end{align}
Here, the matrix elements $T_{ij}$ in the second line of $\hat T$ are given by
\begin{align}
T_{01}&=yz'+zy',
\\
T_{11}&=xx'+xz'+yy'+zx'+zz',
\\
T_{21}&=xy'+yx'+yz'+zy',
\\
T_{31}&=xy'+yz'+zy',
\\
T_{41}&=yx'+yz'+zy',
\\
T_{51}&=xz'+yy'+zz',
\\
T_{61}&=yy'+zx'+zz',
\\
T_{71}&=xz'+yy'+zx'+zz',
\end{align}
$T_{ij}$ in the third line are given by
\begin{align}
T_{02}&=yy'+zz',
\\
T_{12}&=xy'+yx'+yz'+zy',
\\
T_{22}&=xx'+xz'+yy'+zx'+zz',
\\
T_{32}&=xz'+yy'+zz',
\\
T_{42}&=yy'+zx'+zz',
\\
T_{52}&=xy'+yz'+zy',
\\
T_{62}&=yx'+yz'+zy',
\\
T_{72}&=xy'+yx'+yz'+zy',
\end{align}
and, the others are given by
\begin{align}
T_{33}&=xx'+zx',
\\
T_{44}&=xx'+xz',
\\
T_{55}&= xx'+zx',
\\
T_{66}&=xx'+xz'.
\end{align}

By solving these eight self-consistent equations [Eq.~(\ref{vertex})], $J_0$--$J_7$ are determined.
Substituting them into Eq.~(\ref{conductivity}), the conductivity can be rewritten as
\begin{widetext}
\begin{align}
&&\sigma(\epsilon)= \frac{2\hbar e^2 v^2}{3}\int_0^\infty \frac{k'^2dk'}{(2\pi)^3}\mbox{Re}\left[-\frac{J_0^{++}+J_1^{++}+J_2^{++}+J_3^{++}+J_4^{++}+J_5^{++}+J_6^{++}+J_7^{++}}{(X+Y+Z)^2}\right.\nonumber\\
&&-\frac{J_0^{++}-J_1^{++}+J_2^{++}+J_3^{++}+J_4^{++}-J_5^{++}-J_6^{++}-J_7^{++}}{(X-Y+Z)^2}-\frac{2J_0^{++}+J_3^{++}+J_4^{++}+J_5^{++}+J_6^{++}}{X(X+Y+Z)}\nonumber\\
&&-\frac{2J_0^{++}+J_3^{++}+J_4^{++}-J_5^{++}-J_6^{++}}{X(X-Y+Z)}+\frac{J_0^{+-}+J_1^{+-}+J_2^{+-}+J_3^{+-}+J_4^{+-}+J_5^{+-}+J_6^{+-}+J_7^{+-}}{|X+Y+Z|^2}\nonumber\\
&&+\frac{J_0^{+-}-J_1^{+-}+J_2^{+-}+J_3^{+-}+J_4^{+-}-J_5^{+-}-J_6^{+-}-J_7^{+-}}{|X-Y+Z|^2}+\frac{J_0^{+-}+J_4^{+-}+J_6^{+-}}{X(X^*+Y^*+Z^*)}\nonumber\\
&&+\frac{J_0^{+-}+J_4^{+-}-J_6^{+-}}{X(X^*-Y^*+Z^*)}\left.+\frac{J_0^{+-}+J_3^{+-}+J_5^{+-}}{X^*(X+Y+Z)}+\frac{J_0^{+-}+J_3^{+-}-J_5^{+-}}{X^*(X-Y+Z)}\right],
\end{align}
\end{widetext}
where $J_i^{ss'}=J_i(k',\epsilon+is0,\epsilon+is'0)$, $X=X(k',\epsilon+i0)$, and so on.

\section{Useful relations}\label{integral}
Consider $\bm{n}_{\perp 1}$, $\bm{n}_{\perp 2}$, and $\bm{n}$ as three mutually perpendicular unit vectors in three-dimensional space. Let $S_x$, $S_y$, and $S_z$ represent the $3 \times 3$ spin-1 representation matrices introduced in the primary text. The following valuable relationships can be derived from these parameters.
\begin{align}
	&
(\hat{\bm{S}}\cdot\bm{n})^3=(\hat{\bm{S}}\cdot\bm{n}),
\label{UR1}
\\&
(\hat{\bm{S}}\cdot\bm{n})^2\hat{S}_{i}(\hat{\bm{S}}\cdot\bm{n})^2=(\hat{\bm{S}}\cdot\bm{n})\hat{S}_{i}(\hat{\bm{S}}\cdot\bm{n})=n_{i}(\hat{\bm{S}}\cdot\bm{n}),
\\&
(\hat{\bm{S}}\cdot\bm{n}_{\perp 1})^2+(\hat{\bm{S}}\cdot\bm{n}_{\perp 2})^2+(\hat{\bm{S}}\cdot\bm{n})^2=2\hat{S}_0,
\\&
(\hat{\bm{S}}\cdot\bm{n}_{\perp 1})\hat{S}_{i}(\hat{\bm{S}}\cdot\bm{n}_{\perp 1})+(\hat{\bm{S}}\cdot\bm{n}_{\perp 2})\hat{S}_{i}(\hat{\bm{S}}\cdot\bm{n}_{\perp 2})
\nonumber\\&
+(\hat{\bm{S}}\cdot\bm{n})\hat{S}_{i}(\hat{\bm{S}}\cdot\bm{n})=\hat{S}_{i},
\\&
(\hat{\bm{S}}\cdot\bm{n}_{\perp 1})\hat{S}_{i}\hat{S}_{j}(\hat{\bm{S}}\cdot\bm{n}_{\perp 1})+(\hat{\bm{S}}\cdot\bm{n}_{\perp 2})\hat{S}_{i}\hat{S}_{j}(\hat{\bm{S}}\cdot\bm{n}_{\perp 2})
\nonumber\\&
+(\hat{\bm{S}}\cdot\bm{n})\hat{S}_{i}\hat{S}_{j}(\hat{\bm{S}}\cdot\bm{n})=-\hat{S}_{j}\hat{S}_{i}+2\delta_{ij}\hat{S}_0.
\label{UR5}
\end{align}

An arbitrary unit vector $\boldsymbol{n}'$ is written as
\begin{align}
	\bm{n'}
	=\bm{n}_{\perp 1}\sin\theta\cos\phi+\bm{n}_{\perp 2}\sin\theta\sin\phi+\bm{n}\cos\theta,
\end{align}
where $\theta$ represents the angle between $\bm{n}$ and $\bm{n'}$, while $\phi$ signifies the azimuth angle within the $\bm{n}_{\perp1}$-$\bm{n}_{\perp 2}$ planes.

By Eqs.~(\ref{UR1})--(\ref{UR5}), we derive the following equations as
\begin{align}
	&
\int_0^{2\pi}\int_0^\pi d\theta d\phi|u(\bm{k}-\bm{k'})|^2(\hat{\bm{S}}\cdot\bm{n'})
=(\hat{\bm{S}}\cdot\bm{n})V_1^2(k,k'),\label{int1}
\\&
\int_0^{2\pi}\int_0^\pi d\theta d\phi|u(\bm{k}-\bm{k'})|^2(\hat{\bm{S}}\cdot\bm{n'})^2
\nonumber\\&
=\left(\frac{3}{2}V_2^2(k,k')-\frac{1}{2}V_0^2(k,k')\right)(\hat{\bm{S}}\cdot\bm{n})^2
\nonumber\\&\quad
+\left(V_0^2(k,k')-V_2^2(k,k')\right)\hat{S}_0,
\\&
\int_0^{2\pi}\int_0^\pi d\theta d\phi|u(\bm{k}-\bm{k'})|^2n'_x\hat{S}_0
=V_1^2(k,k')n_x\hat{S}_0,
\\&
\int_0^{2\pi}\int_0^\pi d\theta d\phi|u(\bm{k}-\bm{k'})|^2n'_x(\hat{\bm{S}}\cdot\bm{n'})
\nonumber\\&
=\left(\frac{3}{2}V_2^2(k,k')-\frac{1}{2}V_0^2(k,k')\right)n_x(\hat{\bm{S}}\cdot\bm{n})
\nonumber\\&\quad
+\frac{1}{2}(V_0^2(k,k')-V_2^2(k,k'))\hat{S}_{x},
\\&
\int_0^{2\pi}\int_0^\pi d\theta d\phi|u(\bm{k}-\bm{k'})|^2n'_x(\hat{\bm{S}}\cdot\bm{n'})^2
\nonumber\\&
=\frac{1}{2}(V_1^2(k,k')-V_3^2(k,k'))\hat{S}_{x}(\hat{\bm{S}}\cdot\bm{n})
\nonumber\\&\quad
+\frac{1}{2}(V_1^2(k,k')-V_3^2(k,k'))(\hat{\bm{S}}\cdot\bm{n})\hat{S}_{x}
\nonumber\\&\quad
+(V_1^2(k,k')-V_3^2(k,k'))n_x\hat{S}_0
\nonumber\\&\quad
+\left(\frac{5}{2}V_3^2(k,k')-\frac{3}{2}V_1^2(k,k')\right)n_x(\hat{\bm{S}}\cdot\bm{n})^2.\label{int2}
\end{align}

\section{Gauss potential}\label{appendix:gauss}
\begin{figure}
\includegraphics[width=7cm]{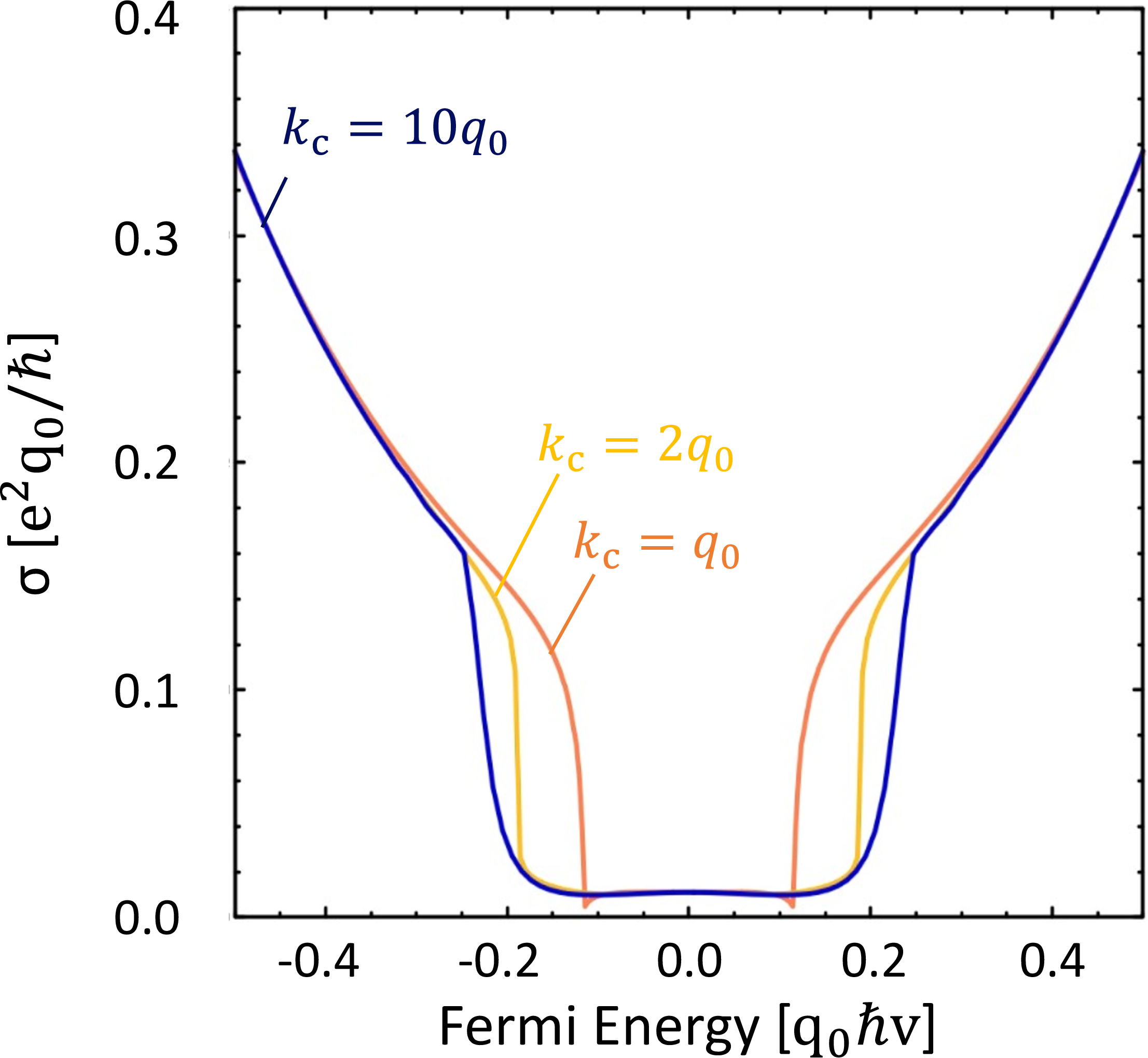}
\caption{(Color online) The conductivity derived by the SCBA for $k_{\text{c}}=q_0$ (orange line), $k_{\text{c}}=2q_0$ (yellow line) and $k_{\text{c}}=10q_0$ (blue line) under the Gaussian potential ($W=2$). 
}
\label{gauss}
\end{figure}
Here we show the results for the cutoff wavenumber dependence of conductivity under the Gaussian potential \cite{Kikuchi2022}.

The Gaussian potential is defined by
\begin{eqnarray}
\label{gauss2}
U(\bm{r}) &=&  \frac{\pm u_0}{(\sqrt{\pi}d_0)^3}\exp(-\frac{r^2}{d_0^2}),
\end{eqnarray}
where $d_0$ is the characteristic length scale and $\pm u_0$ is the strength of the impurity potential. 
The Fourier transforms are obtained to be
\begin{align}
 u(\bm{k}) =  \pm u_0\exp(-\frac{k^2}{q_0^2}),
\end{align}
with $q_0 = 2/d_0$. 
Note that the definition of $q_0$ differs from that of the Coulomb potential case (Eq.~(\ref{coulombq_0})).
We also define a parameter characterizing the scattering strength
\begin{eqnarray}
W = q_0n_{\text{i}}u_0^2,
\end{eqnarray}
where $n_{\text{i}}$ is the number of scatterers per unit volume.

Numerical calculations by the SCBA yield the conductivity as shown in Fig.~\ref{gauss}. In the low-energy region, the conductivity is suppressed by the flat band with zero group velocity, and  nearly independent of the cutoff wavenumber $k_{\text{c}}$.

\bibliography{ref}

\begin{thebibliography}{52}%
\makeatletter
\providecommand \@ifxundefined [1]{%
 \@ifx{#1\undefined}
}%
\providecommand \@ifnum [1]{%
 \ifnum #1\expandafter \@firstoftwo
 \else \expandafter \@secondoftwo
 \fi
}%
\providecommand \@ifx [1]{%
 \ifx #1\expandafter \@firstoftwo
 \else \expandafter \@secondoftwo
 \fi
}%
\providecommand \natexlab [1]{#1}%
\providecommand \enquote  [1]{``#1''}%
\providecommand \bibnamefont  [1]{#1}%
\providecommand \bibfnamefont [1]{#1}%
\providecommand \citenamefont [1]{#1}%
\providecommand \href@noop [0]{\@secondoftwo}%
\providecommand \href [0]{\begingroup \@sanitize@url \@href}%
\providecommand \@href[1]{\@@startlink{#1}\@@href}%
\providecommand \@@href[1]{\endgroup#1\@@endlink}%
\providecommand \@sanitize@url [0]{\catcode `\\12\catcode `\$12\catcode
  `\&12\catcode `\#12\catcode `\^12\catcode `\_12\catcode `\%12\relax}%
\providecommand \@@startlink[1]{}%
\providecommand \@@endlink[0]{}%
\providecommand \url  [0]{\begingroup\@sanitize@url \@url }%
\providecommand \@url [1]{\endgroup\@href {#1}{\urlprefix }}%
\providecommand \urlprefix  [0]{URL }%
\providecommand \Eprint [0]{\href }%
\providecommand \doibase [0]{https://doi.org/}%
\providecommand \selectlanguage [0]{\@gobble}%
\providecommand \bibinfo  [0]{\@secondoftwo}%
\providecommand \bibfield  [0]{\@secondoftwo}%
\providecommand \translation [1]{[#1]}%
\providecommand \BibitemOpen [0]{}%
\providecommand \bibitemStop [0]{}%
\providecommand \bibitemNoStop [0]{.\EOS\space}%
\providecommand \EOS [0]{\spacefactor3000\relax}%
\providecommand \BibitemShut  [1]{\csname bibitem#1\endcsname}%
\let\auto@bib@innerbib\@empty
\bibitem [{\citenamefont {Ma\~nes}(2012)}]{Ma2012}%
  \BibitemOpen
  \bibfield  {author} {\bibinfo {author} {\bibfnamefont {J.~L.}\ \bibnamefont
  {Ma\~nes}},\ }\bibfield  {title} {\bibinfo {title} {Existence of bulk chiral
  fermions and crystal symmetry},\ }\href
  {https://doi.org/10.1103/PhysRevB.85.155118} {\bibfield  {journal} {\bibinfo
  {journal} {Phys. Rev. B}\ }\textbf {\bibinfo {volume} {85}},\ \bibinfo
  {pages} {155118} (\bibinfo {year} {2012})}\BibitemShut {NoStop}%
\bibitem [{\citenamefont {Bradlyn}\ \emph {et~al.}(2016)\citenamefont
  {Bradlyn}, \citenamefont {Cano}, \citenamefont {Wang}, \citenamefont
  {Vergniory}, \citenamefont {Felser}, \citenamefont {Cava},\ and\
  \citenamefont {Bernevig}}]{Beyond2016}%
  \BibitemOpen
  \bibfield  {author} {\bibinfo {author} {\bibfnamefont {B.}~\bibnamefont
  {Bradlyn}}, \bibinfo {author} {\bibfnamefont {J.}~\bibnamefont {Cano}},
  \bibinfo {author} {\bibfnamefont {Z.}~\bibnamefont {Wang}}, \bibinfo {author}
  {\bibfnamefont {M.~G.}\ \bibnamefont {Vergniory}}, \bibinfo {author}
  {\bibfnamefont {C.}~\bibnamefont {Felser}}, \bibinfo {author} {\bibfnamefont
  {R.~J.}\ \bibnamefont {Cava}},\ and\ \bibinfo {author} {\bibfnamefont
  {B.~A.}\ \bibnamefont {Bernevig}},\ }\bibfield  {title} {\bibinfo {title}
  {Beyond dirac and weyl fermions: Unconventional quasiparticles in
  conventional crystals},\ }\href {http://dx.doi.org/10.1126/science.aaf5037}
  {\bibfield  {journal} {\bibinfo  {journal} {Science}\ }\textbf {\bibinfo
  {volume} {353}},\ \bibinfo {pages} {aaf5037} (\bibinfo {year}
  {2016})}\BibitemShut {NoStop}%
\bibitem [{\citenamefont {Tang}\ \emph {et~al.}(2017)\citenamefont {Tang},
  \citenamefont {Zhou},\ and\ \citenamefont {Zhang}}]{Tang2017-kk}%
  \BibitemOpen
  \bibfield  {author} {\bibinfo {author} {\bibfnamefont {P.}~\bibnamefont
  {Tang}}, \bibinfo {author} {\bibfnamefont {Q.}~\bibnamefont {Zhou}},\ and\
  \bibinfo {author} {\bibfnamefont {S.-C.}\ \bibnamefont {Zhang}},\ }\bibfield
  {title} {\bibinfo {title} {{Multiple Types of Topological Fermions in
  Transition Metal Silicides}},\ }\href
  {https://doi.org/10.1103/PhysRevLett.119.206402} {\bibfield  {journal}
  {\bibinfo  {journal} {Phys. Rev. Lett.}\ }\textbf {\bibinfo {volume} {119}},\
  \bibinfo {pages} {206402} (\bibinfo {year} {2017})}\BibitemShut {NoStop}%
\bibitem [{\citenamefont {Chang}\ \emph {et~al.}(2017)\citenamefont {Chang},
  \citenamefont {Xu}, \citenamefont {Wieder}, \citenamefont {Sanchez},
  \citenamefont {Huang}, \citenamefont {Belopolski}, \citenamefont {Chang},
  \citenamefont {Zhang}, \citenamefont {Bansil}, \citenamefont {Lin},\ and\
  \citenamefont {Hasan}}]{Chang2017}%
  \BibitemOpen
  \bibfield  {author} {\bibinfo {author} {\bibfnamefont {G.}~\bibnamefont
  {Chang}}, \bibinfo {author} {\bibfnamefont {S.-Y.}\ \bibnamefont {Xu}},
  \bibinfo {author} {\bibfnamefont {B.~J.}\ \bibnamefont {Wieder}}, \bibinfo
  {author} {\bibfnamefont {D.~S.}\ \bibnamefont {Sanchez}}, \bibinfo {author}
  {\bibfnamefont {S.-M.}\ \bibnamefont {Huang}}, \bibinfo {author}
  {\bibfnamefont {I.}~\bibnamefont {Belopolski}}, \bibinfo {author}
  {\bibfnamefont {T.-R.}\ \bibnamefont {Chang}}, \bibinfo {author}
  {\bibfnamefont {S.}~\bibnamefont {Zhang}}, \bibinfo {author} {\bibfnamefont
  {A.}~\bibnamefont {Bansil}}, \bibinfo {author} {\bibfnamefont
  {H.}~\bibnamefont {Lin}},\ and\ \bibinfo {author} {\bibfnamefont {M.~Z.}\
  \bibnamefont {Hasan}},\ }\bibfield  {title} {\bibinfo {title}
  {{Unconventional Chiral Fermions and Large Topological Fermi Arcs in RhSi}},\
  }\href {https://doi.org/10.1103/PhysRevLett.119.206401} {\bibfield  {journal}
  {\bibinfo  {journal} {Phys. Rev. Lett.}\ }\textbf {\bibinfo {volume} {119}},\
  \bibinfo {pages} {206401} (\bibinfo {year} {2017})}\BibitemShut {NoStop}%
\bibitem [{\citenamefont {Pshenay-Severin}\ \emph
  {et~al.}(2018{\natexlab{a}})\citenamefont {Pshenay-Severin}, \citenamefont
  {Ivanov}, \citenamefont {Burkov},\ and\ \citenamefont
  {Burkov}}]{Pshenay-Severin_2018}%
  \BibitemOpen
  \bibfield  {author} {\bibinfo {author} {\bibfnamefont {D.~A.}\ \bibnamefont
  {Pshenay-Severin}}, \bibinfo {author} {\bibfnamefont {Y.~V.}\ \bibnamefont
  {Ivanov}}, \bibinfo {author} {\bibfnamefont {A.~A.}\ \bibnamefont {Burkov}},\
  and\ \bibinfo {author} {\bibfnamefont {A.~T.}\ \bibnamefont {Burkov}},\
  }\bibfield  {title} {\bibinfo {title} {{Band structure and unconventional
  electronic topology of CoSi}},\ }\href
  {https://doi.org/10.1088/1361-648X/aab0ba} {\bibfield  {journal} {\bibinfo
  {journal} {J. Phys.: Condens. Matter}\ }\textbf {\bibinfo {volume} {30}},\
  \bibinfo {pages} {135501} (\bibinfo {year} {2018}{\natexlab{a}})}\BibitemShut
  {NoStop}%
\bibitem [{\citenamefont {Takane}\ \emph {et~al.}(2019)\citenamefont {Takane},
  \citenamefont {Wang}, \citenamefont {Souma}, \citenamefont {Nakayama},
  \citenamefont {Nakamura}, \citenamefont {Oinuma}, \citenamefont {Nakata},
  \citenamefont {Iwasawa}, \citenamefont {Cacho}, \citenamefont {Kim},
  \citenamefont {Horiba}, \citenamefont {Kumigashira}, \citenamefont
  {Takahashi}, \citenamefont {Ando},\ and\ \citenamefont {Sato}}]{Takane2019}%
  \BibitemOpen
  \bibfield  {author} {\bibinfo {author} {\bibfnamefont {D.}~\bibnamefont
  {Takane}}, \bibinfo {author} {\bibfnamefont {Z.}~\bibnamefont {Wang}},
  \bibinfo {author} {\bibfnamefont {S.}~\bibnamefont {Souma}}, \bibinfo
  {author} {\bibfnamefont {K.}~\bibnamefont {Nakayama}}, \bibinfo {author}
  {\bibfnamefont {T.}~\bibnamefont {Nakamura}}, \bibinfo {author}
  {\bibfnamefont {H.}~\bibnamefont {Oinuma}}, \bibinfo {author} {\bibfnamefont
  {Y.}~\bibnamefont {Nakata}}, \bibinfo {author} {\bibfnamefont
  {H.}~\bibnamefont {Iwasawa}}, \bibinfo {author} {\bibfnamefont
  {C.}~\bibnamefont {Cacho}}, \bibinfo {author} {\bibfnamefont
  {T.}~\bibnamefont {Kim}}, \bibinfo {author} {\bibfnamefont {K.}~\bibnamefont
  {Horiba}}, \bibinfo {author} {\bibfnamefont {H.}~\bibnamefont {Kumigashira}},
  \bibinfo {author} {\bibfnamefont {T.}~\bibnamefont {Takahashi}}, \bibinfo
  {author} {\bibfnamefont {Y.}~\bibnamefont {Ando}},\ and\ \bibinfo {author}
  {\bibfnamefont {T.}~\bibnamefont {Sato}},\ }\bibfield  {title} {\bibinfo
  {title} {{Observation of Chiral Fermions with a Large Topological Charge and
  Associated Fermi-Arc Surface States in CoSi}},\ }\href
  {https://doi.org/10.1103/PhysRevLett.122.076402} {\bibfield  {journal}
  {\bibinfo  {journal} {Phys. Rev. Lett.}\ }\textbf {\bibinfo {volume} {122}},\
  \bibinfo {pages} {076402} (\bibinfo {year} {2019})}\BibitemShut {NoStop}%
\bibitem [{\citenamefont {Rao}\ \emph {et~al.}(2019)\citenamefont {Rao},
  \citenamefont {Li}, \citenamefont {Zhang}, \citenamefont {Tian},
  \citenamefont {Li}, \citenamefont {Fu}, \citenamefont {Tang}, \citenamefont
  {Wang}, \citenamefont {Li}, \citenamefont {Fan}, \citenamefont {Li},
  \citenamefont {Huang}, \citenamefont {Liu}, \citenamefont {Long},
  \citenamefont {Fang}, \citenamefont {Weng}, \citenamefont {Shi},
  \citenamefont {Lei}, \citenamefont {Sun}, \citenamefont {Qian},\ and\
  \citenamefont {Ding}}]{Rao2019-ts}%
  \BibitemOpen
  \bibfield  {author} {\bibinfo {author} {\bibfnamefont {Z.}~\bibnamefont
  {Rao}}, \bibinfo {author} {\bibfnamefont {H.}~\bibnamefont {Li}}, \bibinfo
  {author} {\bibfnamefont {T.}~\bibnamefont {Zhang}}, \bibinfo {author}
  {\bibfnamefont {S.}~\bibnamefont {Tian}}, \bibinfo {author} {\bibfnamefont
  {C.}~\bibnamefont {Li}}, \bibinfo {author} {\bibfnamefont {B.}~\bibnamefont
  {Fu}}, \bibinfo {author} {\bibfnamefont {C.}~\bibnamefont {Tang}}, \bibinfo
  {author} {\bibfnamefont {L.}~\bibnamefont {Wang}}, \bibinfo {author}
  {\bibfnamefont {Z.}~\bibnamefont {Li}}, \bibinfo {author} {\bibfnamefont
  {W.}~\bibnamefont {Fan}}, \bibinfo {author} {\bibfnamefont {J.}~\bibnamefont
  {Li}}, \bibinfo {author} {\bibfnamefont {Y.}~\bibnamefont {Huang}}, \bibinfo
  {author} {\bibfnamefont {Z.}~\bibnamefont {Liu}}, \bibinfo {author}
  {\bibfnamefont {Y.}~\bibnamefont {Long}}, \bibinfo {author} {\bibfnamefont
  {C.}~\bibnamefont {Fang}}, \bibinfo {author} {\bibfnamefont {H.}~\bibnamefont
  {Weng}}, \bibinfo {author} {\bibfnamefont {Y.}~\bibnamefont {Shi}}, \bibinfo
  {author} {\bibfnamefont {H.}~\bibnamefont {Lei}}, \bibinfo {author}
  {\bibfnamefont {Y.}~\bibnamefont {Sun}}, \bibinfo {author} {\bibfnamefont
  {T.}~\bibnamefont {Qian}},\ and\ \bibinfo {author} {\bibfnamefont
  {H.}~\bibnamefont {Ding}},\ }\bibfield  {title} {\bibinfo {title}
  {{Observation of unconventional chiral fermions with long Fermi arcs in
  CoSi}},\ }\href {https://doi.org/10.1038/s41586-019-1031-8} {\bibfield
  {journal} {\bibinfo  {journal} {Nature}\ }\textbf {\bibinfo {volume} {567}},\
  \bibinfo {pages} {496} (\bibinfo {year} {2019})}\BibitemShut {NoStop}%
\bibitem [{\citenamefont {Sanchez}\ \emph {et~al.}(2019)\citenamefont
  {Sanchez}, \citenamefont {Belopolski}, \citenamefont {Cochran}, \citenamefont
  {Xu}, \citenamefont {Yin}, \citenamefont {Chang}, \citenamefont {Xie},
  \citenamefont {Manna}, \citenamefont {S{\"u}{\ss}}, \citenamefont {Huang},
  \citenamefont {Alidoust}, \citenamefont {Multer}, \citenamefont {Zhang},
  \citenamefont {Shumiya}, \citenamefont {Wang}, \citenamefont {Wang},
  \citenamefont {Chang}, \citenamefont {Felser}, \citenamefont {Xu},
  \citenamefont {Jia}, \citenamefont {Lin},\ and\ \citenamefont
  {Hasan}}]{Sanchez2019-by}%
  \BibitemOpen
  \bibfield  {author} {\bibinfo {author} {\bibfnamefont {D.~S.}\ \bibnamefont
  {Sanchez}}, \bibinfo {author} {\bibfnamefont {I.}~\bibnamefont {Belopolski}},
  \bibinfo {author} {\bibfnamefont {T.~A.}\ \bibnamefont {Cochran}}, \bibinfo
  {author} {\bibfnamefont {X.}~\bibnamefont {Xu}}, \bibinfo {author}
  {\bibfnamefont {J.-X.}\ \bibnamefont {Yin}}, \bibinfo {author} {\bibfnamefont
  {G.}~\bibnamefont {Chang}}, \bibinfo {author} {\bibfnamefont
  {W.}~\bibnamefont {Xie}}, \bibinfo {author} {\bibfnamefont {K.}~\bibnamefont
  {Manna}}, \bibinfo {author} {\bibfnamefont {V.}~\bibnamefont {S{\"u}{\ss}}},
  \bibinfo {author} {\bibfnamefont {C.-Y.}\ \bibnamefont {Huang}}, \bibinfo
  {author} {\bibfnamefont {N.}~\bibnamefont {Alidoust}}, \bibinfo {author}
  {\bibfnamefont {D.}~\bibnamefont {Multer}}, \bibinfo {author} {\bibfnamefont
  {S.~S.}\ \bibnamefont {Zhang}}, \bibinfo {author} {\bibfnamefont
  {N.}~\bibnamefont {Shumiya}}, \bibinfo {author} {\bibfnamefont
  {X.}~\bibnamefont {Wang}}, \bibinfo {author} {\bibfnamefont {G.-Q.}\
  \bibnamefont {Wang}}, \bibinfo {author} {\bibfnamefont {T.-R.}\ \bibnamefont
  {Chang}}, \bibinfo {author} {\bibfnamefont {C.}~\bibnamefont {Felser}},
  \bibinfo {author} {\bibfnamefont {S.-Y.}\ \bibnamefont {Xu}}, \bibinfo
  {author} {\bibfnamefont {S.}~\bibnamefont {Jia}}, \bibinfo {author}
  {\bibfnamefont {H.}~\bibnamefont {Lin}},\ and\ \bibinfo {author}
  {\bibfnamefont {M.~Z.}\ \bibnamefont {Hasan}},\ }\bibfield  {title} {\bibinfo
  {title} {{Topological chiral crystals with helicoid-arc quantum states}},\
  }\href {https://doi.org/10.1038/s41586-019-1037-2} {\bibfield  {journal}
  {\bibinfo  {journal} {Nature}\ }\textbf {\bibinfo {volume} {567}},\ \bibinfo
  {pages} {500} (\bibinfo {year} {2019})}\BibitemShut {NoStop}%
\bibitem [{\citenamefont {Mozaffari}\ \emph {et~al.}(2020)\citenamefont
  {Mozaffari}, \citenamefont {Aryal}, \citenamefont {Sch\"onemann},
  \citenamefont {Chen}, \citenamefont {Zheng}, \citenamefont {McCandless},
  \citenamefont {Chan}, \citenamefont {Manousakis},\ and\ \citenamefont
  {Balicas}}]{Mozaffari2020}%
  \BibitemOpen
  \bibfield  {author} {\bibinfo {author} {\bibfnamefont {S.}~\bibnamefont
  {Mozaffari}}, \bibinfo {author} {\bibfnamefont {N.}~\bibnamefont {Aryal}},
  \bibinfo {author} {\bibfnamefont {R.}~\bibnamefont {Sch\"onemann}}, \bibinfo
  {author} {\bibfnamefont {K.-W.}\ \bibnamefont {Chen}}, \bibinfo {author}
  {\bibfnamefont {W.}~\bibnamefont {Zheng}}, \bibinfo {author} {\bibfnamefont
  {G.~T.}\ \bibnamefont {McCandless}}, \bibinfo {author} {\bibfnamefont
  {J.~Y.}\ \bibnamefont {Chan}}, \bibinfo {author} {\bibfnamefont
  {E.}~\bibnamefont {Manousakis}},\ and\ \bibinfo {author} {\bibfnamefont
  {L.}~\bibnamefont {Balicas}},\ }\bibfield  {title} {\bibinfo {title}
  {{Multiple Dirac nodes and symmetry protected Dirac nodal line in
  orthorhombic $\ensuremath{\alpha}$-RhSi}},\ }\href
  {https://doi.org/10.1103/PhysRevB.102.115131} {\bibfield  {journal} {\bibinfo
   {journal} {Phys. Rev. B}\ }\textbf {\bibinfo {volume} {102}},\ \bibinfo
  {pages} {115131} (\bibinfo {year} {2020})}\BibitemShut {NoStop}%
\bibitem [{\citenamefont {Li}\ \emph {et~al.}(2019{\natexlab{a}})\citenamefont
  {Li}, \citenamefont {Xu}, \citenamefont {Rao}, \citenamefont {Zhou},
  \citenamefont {Wang}, \citenamefont {Zhou}, \citenamefont {Tian},
  \citenamefont {Gao}, \citenamefont {Li}, \citenamefont {Huang}, \citenamefont
  {Lei}, \citenamefont {Weng}, \citenamefont {Sun}, \citenamefont {Xia},
  \citenamefont {Qian},\ and\ \citenamefont {Ding}}]{Li2019_RhSn}%
  \BibitemOpen
  \bibfield  {author} {\bibinfo {author} {\bibfnamefont {H.}~\bibnamefont
  {Li}}, \bibinfo {author} {\bibfnamefont {S.}~\bibnamefont {Xu}}, \bibinfo
  {author} {\bibfnamefont {Z.-C.}\ \bibnamefont {Rao}}, \bibinfo {author}
  {\bibfnamefont {L.-Q.}\ \bibnamefont {Zhou}}, \bibinfo {author}
  {\bibfnamefont {Z.-J.}\ \bibnamefont {Wang}}, \bibinfo {author}
  {\bibfnamefont {S.-M.}\ \bibnamefont {Zhou}}, \bibinfo {author}
  {\bibfnamefont {S.-J.}\ \bibnamefont {Tian}}, \bibinfo {author}
  {\bibfnamefont {S.-Y.}\ \bibnamefont {Gao}}, \bibinfo {author} {\bibfnamefont
  {J.-J.}\ \bibnamefont {Li}}, \bibinfo {author} {\bibfnamefont {Y.-B.}\
  \bibnamefont {Huang}}, \bibinfo {author} {\bibfnamefont {H.-C.}\ \bibnamefont
  {Lei}}, \bibinfo {author} {\bibfnamefont {H.-M.}\ \bibnamefont {Weng}},
  \bibinfo {author} {\bibfnamefont {Y.-J.}\ \bibnamefont {Sun}}, \bibinfo
  {author} {\bibfnamefont {T.-L.}\ \bibnamefont {Xia}}, \bibinfo {author}
  {\bibfnamefont {T.}~\bibnamefont {Qian}},\ and\ \bibinfo {author}
  {\bibfnamefont {H.}~\bibnamefont {Ding}},\ }\bibfield  {title} {\bibinfo
  {title} {Chiral fermion reversal in chiral crystals},\ }\href
  {https://doi.org/10.1038/s41467-019-13435-4} {\bibfield  {journal} {\bibinfo
  {journal} {Nat. Commun.}\ }\textbf {\bibinfo {volume} {10}},\ \bibinfo
  {pages} {5505} (\bibinfo {year} {2019}{\natexlab{a}})}\BibitemShut {NoStop}%
\bibitem [{\citenamefont {Schröter}\ \emph {et~al.}(2019)\citenamefont
  {Schröter}, \citenamefont {Pei}, \citenamefont {Vergniory}, \citenamefont
  {Manna}, \citenamefont {de~Juan}, \citenamefont {Krieger}, \citenamefont
  {Süss}, \citenamefont {Schmidt}, \citenamefont {Bradlyn}, \citenamefont
  {Kim}, \citenamefont {Schmitt}, \citenamefont {Cacho}, \citenamefont
  {Felser}, \citenamefont {Strocov},\ and\ \citenamefont {Chen}}]{Schr2019}%
  \BibitemOpen
  \bibfield  {author} {\bibinfo {author} {\bibfnamefont {N.~B.~M.}\
  \bibnamefont {Schröter}}, \bibinfo {author} {\bibfnamefont {D.}~\bibnamefont
  {Pei}}, \bibinfo {author} {\bibfnamefont {Y.}~\bibnamefont {Vergniory},
  \bibfnamefont {Maia G.~Sun}}, \bibinfo {author} {\bibfnamefont
  {K.}~\bibnamefont {Manna}}, \bibinfo {author} {\bibfnamefont
  {F.}~\bibnamefont {de~Juan}}, \bibinfo {author} {\bibfnamefont {J.~A.}\
  \bibnamefont {Krieger}}, \bibinfo {author} {\bibfnamefont {V.}~\bibnamefont
  {Süss}}, \bibinfo {author} {\bibfnamefont {P.}~\bibnamefont {Schmidt},
  \bibfnamefont {Marcus~Dudin}}, \bibinfo {author} {\bibfnamefont
  {B.}~\bibnamefont {Bradlyn}}, \bibinfo {author} {\bibfnamefont {T.~K.}\
  \bibnamefont {Kim}}, \bibinfo {author} {\bibfnamefont {T.}~\bibnamefont
  {Schmitt}}, \bibinfo {author} {\bibfnamefont {C.}~\bibnamefont {Cacho}},
  \bibinfo {author} {\bibfnamefont {C.}~\bibnamefont {Felser}}, \bibinfo
  {author} {\bibfnamefont {V.~N.}\ \bibnamefont {Strocov}},\ and\ \bibinfo
  {author} {\bibfnamefont {Y.}~\bibnamefont {Chen}},\ }\bibfield  {title}
  {\bibinfo {title} {{Chiral topological semimetal with multifold band
  crossings and long Fermi arcs}},\ }\href
  {https://doi.org/10.1038/s41567-019-0511-y} {\bibfield  {journal} {\bibinfo
  {journal} {Nat. Phys.}\ }\textbf {\bibinfo {volume} {15}},\ \bibinfo {pages}
  {759–765} (\bibinfo {year} {2019})}\BibitemShut {NoStop}%
\bibitem [{\citenamefont {Wu}\ \emph {et~al.}(2019)\citenamefont {Wu},
  \citenamefont {Mi}, \citenamefont {Li}, \citenamefont {Wu}, \citenamefont
  {Li}, \citenamefont {Song}, \citenamefont {Liu}, \citenamefont {Li},\ and\
  \citenamefont {Luo}}]{Wu_2019}%
  \BibitemOpen
  \bibfield  {author} {\bibinfo {author} {\bibfnamefont {D.~S.}\ \bibnamefont
  {Wu}}, \bibinfo {author} {\bibfnamefont {Z.~Y.}\ \bibnamefont {Mi}}, \bibinfo
  {author} {\bibfnamefont {Y.~J.}\ \bibnamefont {Li}}, \bibinfo {author}
  {\bibfnamefont {W.}~\bibnamefont {Wu}}, \bibinfo {author} {\bibfnamefont
  {P.~L.}\ \bibnamefont {Li}}, \bibinfo {author} {\bibfnamefont {Y.~T.}\
  \bibnamefont {Song}}, \bibinfo {author} {\bibfnamefont {G.~T.}\ \bibnamefont
  {Liu}}, \bibinfo {author} {\bibfnamefont {G.}~\bibnamefont {Li}},\ and\
  \bibinfo {author} {\bibfnamefont {J.~L.}\ \bibnamefont {Luo}},\ }\bibfield
  {title} {\bibinfo {title} {{Single Crystal Growth and Magnetoresistivity of
  Topological Semimetal CoSi}},\ }\href
  {https://doi.org/10.1088/0256-307X/36/7/077102} {\bibfield  {journal}
  {\bibinfo  {journal} {Chin. Phys. Lett.}\ }\textbf {\bibinfo {volume} {36}},\
  \bibinfo {pages} {077102} (\bibinfo {year} {2019})}\BibitemShut {NoStop}%
\bibitem [{\citenamefont {Xu}\ \emph {et~al.}(2019)\citenamefont {Xu},
  \citenamefont {Wang}, \citenamefont {Cochran}, \citenamefont {Sanchez},
  \citenamefont {Chang}, \citenamefont {Belopolski}, \citenamefont {Wang},
  \citenamefont {Liu}, \citenamefont {Tien}, \citenamefont {Gui}, \citenamefont
  {Xie}, \citenamefont {Hasan}, \citenamefont {Chang},\ and\ \citenamefont
  {Jia}}]{Xu2019}%
  \BibitemOpen
  \bibfield  {author} {\bibinfo {author} {\bibfnamefont {X.}~\bibnamefont
  {Xu}}, \bibinfo {author} {\bibfnamefont {X.}~\bibnamefont {Wang}}, \bibinfo
  {author} {\bibfnamefont {T.~A.}\ \bibnamefont {Cochran}}, \bibinfo {author}
  {\bibfnamefont {D.~S.}\ \bibnamefont {Sanchez}}, \bibinfo {author}
  {\bibfnamefont {G.}~\bibnamefont {Chang}}, \bibinfo {author} {\bibfnamefont
  {I.}~\bibnamefont {Belopolski}}, \bibinfo {author} {\bibfnamefont
  {G.}~\bibnamefont {Wang}}, \bibinfo {author} {\bibfnamefont {Y.}~\bibnamefont
  {Liu}}, \bibinfo {author} {\bibfnamefont {H.-J.}\ \bibnamefont {Tien}},
  \bibinfo {author} {\bibfnamefont {X.}~\bibnamefont {Gui}}, \bibinfo {author}
  {\bibfnamefont {W.}~\bibnamefont {Xie}}, \bibinfo {author} {\bibfnamefont
  {M.~Z.}\ \bibnamefont {Hasan}}, \bibinfo {author} {\bibfnamefont {T.-R.}\
  \bibnamefont {Chang}},\ and\ \bibinfo {author} {\bibfnamefont
  {S.}~\bibnamefont {Jia}},\ }\bibfield  {title} {\bibinfo {title} {{Crystal
  growth and quantum oscillations in the topological chiral semimetal CoSi}},\
  }\href {https://doi.org/10.1103/PhysRevB.100.045104} {\bibfield  {journal}
  {\bibinfo  {journal} {Phys. Rev. B}\ }\textbf {\bibinfo {volume} {100}},\
  \bibinfo {pages} {045104} (\bibinfo {year} {2019})}\BibitemShut {NoStop}%
\bibitem [{\citenamefont {Dutta}\ and\ \citenamefont
  {Pandey}(2021)}]{Dutta2021}%
  \BibitemOpen
  \bibfield  {author} {\bibinfo {author} {\bibfnamefont {P.}~\bibnamefont
  {Dutta}}\ and\ \bibinfo {author} {\bibfnamefont {S.~K.}\ \bibnamefont
  {Pandey}},\ }\bibfield  {title} {\bibinfo {title} {{Electronic correlation
  effect on nontrivial topological fermions in CoSi}},\ }\href
  {https://doi.org/10.1140/epjb/s10051-021-00091-1} {\bibfield  {journal}
  {\bibinfo  {journal} {Eur. Phys. J. B}\ }\textbf {\bibinfo {volume} {94}},\
  \bibinfo {pages} {81} (\bibinfo {year} {2021})}\BibitemShut {NoStop}%
\bibitem [{\citenamefont {Tang}\ \emph {et~al.}(2021)\citenamefont {Tang},
  \citenamefont {Lau}, \citenamefont {Nawa}, \citenamefont {Wen}, \citenamefont
  {Xiang}, \citenamefont {Sukegawa}, \citenamefont {Seki}, \citenamefont
  {Miura}, \citenamefont {Takanashi},\ and\ \citenamefont {Mitani}}]{Tang21}%
  \BibitemOpen
  \bibfield  {author} {\bibinfo {author} {\bibfnamefont {K.}~\bibnamefont
  {Tang}}, \bibinfo {author} {\bibfnamefont {Y.-C.}\ \bibnamefont {Lau}},
  \bibinfo {author} {\bibfnamefont {K.}~\bibnamefont {Nawa}}, \bibinfo {author}
  {\bibfnamefont {Z.}~\bibnamefont {Wen}}, \bibinfo {author} {\bibfnamefont
  {Q.}~\bibnamefont {Xiang}}, \bibinfo {author} {\bibfnamefont
  {H.}~\bibnamefont {Sukegawa}}, \bibinfo {author} {\bibfnamefont
  {T.}~\bibnamefont {Seki}}, \bibinfo {author} {\bibfnamefont {Y.}~\bibnamefont
  {Miura}}, \bibinfo {author} {\bibfnamefont {K.}~\bibnamefont {Takanashi}},\
  and\ \bibinfo {author} {\bibfnamefont {S.}~\bibnamefont {Mitani}},\
  }\bibfield  {title} {\bibinfo {title} {{Spin Hall effect in a spin-1 chiral
  semimetal}},\ }\href {https://doi.org/10.1103/PhysRevResearch.3.033101}
  {\bibfield  {journal} {\bibinfo  {journal} {Phys. Rev. Research}\ }\textbf
  {\bibinfo {volume} {3}},\ \bibinfo {pages} {033101} (\bibinfo {year}
  {2021})}\BibitemShut {NoStop}%
\bibitem [{\citenamefont {Kikuchi}\ \emph {et~al.}(2022)\citenamefont
  {Kikuchi}, \citenamefont {Funato},\ and\ \citenamefont
  {Yamakage}}]{Kikuchi2022}%
  \BibitemOpen
  \bibfield  {author} {\bibinfo {author} {\bibfnamefont {R.}~\bibnamefont
  {Kikuchi}}, \bibinfo {author} {\bibfnamefont {T.}~\bibnamefont {Funato}},\
  and\ \bibinfo {author} {\bibfnamefont {A.}~\bibnamefont {Yamakage}},\
  }\bibfield  {title} {\bibinfo {title} {Quantum transport of a spin-1 chiral
  fermion},\ }\href {https://doi.org/10.1103/PhysRevB.106.235204} {\bibfield
  {journal} {\bibinfo  {journal} {Phys. Rev. B}\ }\textbf {\bibinfo {volume}
  {106}},\ \bibinfo {pages} {235204} (\bibinfo {year} {2022})}\BibitemShut
  {NoStop}%
\bibitem [{\citenamefont {Petrova}\ \emph {et~al.}(2023)\citenamefont
  {Petrova}, \citenamefont {Sobolevskiy},\ and\ \citenamefont
  {Stishov}}]{Petrova2023}%
  \BibitemOpen
  \bibfield  {author} {\bibinfo {author} {\bibfnamefont {A.~E.}\ \bibnamefont
  {Petrova}}, \bibinfo {author} {\bibfnamefont {O.~A.}\ \bibnamefont
  {Sobolevskiy}},\ and\ \bibinfo {author} {\bibfnamefont {S.~M.}\ \bibnamefont
  {Stishov}},\ }\bibfield  {title} {\bibinfo {title} {{Magnetoresistance and
  Kohler rule in the topological chiral semimetal CoSi}},\ }\href
  {https://doi.org/10.1103/PhysRevB.107.085136} {\bibfield  {journal} {\bibinfo
   {journal} {Phys. Rev. B}\ }\textbf {\bibinfo {volume} {107}},\ \bibinfo
  {pages} {085136} (\bibinfo {year} {2023})}\BibitemShut {NoStop}%
\bibitem [{\citenamefont {Flicker}\ \emph {et~al.}(2018)\citenamefont
  {Flicker}, \citenamefont {de~Juan}, \citenamefont {Bradlyn}, \citenamefont
  {Morimoto}, \citenamefont {Vergniory},\ and\ \citenamefont
  {Grushin}}]{Flicker18}%
  \BibitemOpen
  \bibfield  {author} {\bibinfo {author} {\bibfnamefont {F.}~\bibnamefont
  {Flicker}}, \bibinfo {author} {\bibfnamefont {F.}~\bibnamefont {de~Juan}},
  \bibinfo {author} {\bibfnamefont {B.}~\bibnamefont {Bradlyn}}, \bibinfo
  {author} {\bibfnamefont {T.}~\bibnamefont {Morimoto}}, \bibinfo {author}
  {\bibfnamefont {M.~G.}\ \bibnamefont {Vergniory}},\ and\ \bibinfo {author}
  {\bibfnamefont {A.~G.}\ \bibnamefont {Grushin}},\ }\bibfield  {title}
  {\bibinfo {title} {Chiral optical response of multifold fermions},\ }\href
  {https://doi.org/10.1103/PhysRevB.98.155145} {\bibfield  {journal} {\bibinfo
  {journal} {Phys. Rev. B}\ }\textbf {\bibinfo {volume} {98}},\ \bibinfo
  {pages} {155145} (\bibinfo {year} {2018})}\BibitemShut {NoStop}%
\bibitem [{\citenamefont {S{\'a}nchez-Mart{\'\i}nez}\ \emph
  {et~al.}(2019)\citenamefont {S{\'a}nchez-Mart{\'\i}nez}, \citenamefont
  {de~Juan},\ and\ \citenamefont {Grushin}}]{Sanchez-Martinez2019-ek}%
  \BibitemOpen
  \bibfield  {author} {\bibinfo {author} {\bibfnamefont {M.-{\'A}.}\
  \bibnamefont {S{\'a}nchez-Mart{\'\i}nez}}, \bibinfo {author} {\bibfnamefont
  {F.}~\bibnamefont {de~Juan}},\ and\ \bibinfo {author} {\bibfnamefont {A.~G.}\
  \bibnamefont {Grushin}},\ }\bibfield  {title} {\bibinfo {title} {{Linear
  optical conductivity of chiral multifold fermions}},\ }\href
  {https://doi.org/10.1103/PhysRevB.99.155145} {\bibfield  {journal} {\bibinfo
  {journal} {Phys. Rev. B}\ }\textbf {\bibinfo {volume} {99}},\ \bibinfo
  {pages} {155145} (\bibinfo {year} {2019})}\BibitemShut {NoStop}%
\bibitem [{\citenamefont {Li}\ \emph {et~al.}(2019{\natexlab{b}})\citenamefont
  {Li}, \citenamefont {Iitaka}, \citenamefont {Zeng},\ and\ \citenamefont
  {Su}}]{Li2019}%
  \BibitemOpen
  \bibfield  {author} {\bibinfo {author} {\bibfnamefont {Z.}~\bibnamefont
  {Li}}, \bibinfo {author} {\bibfnamefont {T.}~\bibnamefont {Iitaka}}, \bibinfo
  {author} {\bibfnamefont {H.}~\bibnamefont {Zeng}},\ and\ \bibinfo {author}
  {\bibfnamefont {H.}~\bibnamefont {Su}},\ }\bibfield  {title} {\bibinfo
  {title} {Optical response of the chiral topological semimetal rhsi},\ }\href
  {https://doi.org/10.1103/PhysRevB.100.155201} {\bibfield  {journal} {\bibinfo
   {journal} {Phys. Rev. B}\ }\textbf {\bibinfo {volume} {100}},\ \bibinfo
  {pages} {155201} (\bibinfo {year} {2019}{\natexlab{b}})}\BibitemShut
  {NoStop}%
\bibitem [{\citenamefont {Habe}(2019)}]{Habe2019-ss}%
  \BibitemOpen
  \bibfield  {author} {\bibinfo {author} {\bibfnamefont {T.}~\bibnamefont
  {Habe}},\ }\bibfield  {title} {\bibinfo {title} {{Dynamical conductivity in
  the multiply degenerate point-nodal semimetal CoSi}},\ }\href
  {https://doi.org/10.1103/PhysRevB.100.245131} {\bibfield  {journal} {\bibinfo
   {journal} {Phys. Rev. B}\ }\textbf {\bibinfo {volume} {100}},\ \bibinfo
  {pages} {245131} (\bibinfo {year} {2019})}\BibitemShut {NoStop}%
\bibitem [{\citenamefont {Maulana}\ \emph {et~al.}(2020)\citenamefont
  {Maulana}, \citenamefont {Manna}, \citenamefont {Uykur}, \citenamefont
  {Felser}, \citenamefont {Dressel},\ and\ \citenamefont
  {Pronin}}]{Maulana2020}%
  \BibitemOpen
  \bibfield  {author} {\bibinfo {author} {\bibfnamefont {L.~Z.}\ \bibnamefont
  {Maulana}}, \bibinfo {author} {\bibfnamefont {K.}~\bibnamefont {Manna}},
  \bibinfo {author} {\bibfnamefont {E.}~\bibnamefont {Uykur}}, \bibinfo
  {author} {\bibfnamefont {C.}~\bibnamefont {Felser}}, \bibinfo {author}
  {\bibfnamefont {M.}~\bibnamefont {Dressel}},\ and\ \bibinfo {author}
  {\bibfnamefont {A.~V.}\ \bibnamefont {Pronin}},\ }\bibfield  {title}
  {\bibinfo {title} {Optical conductivity of multifold fermions: The case of
  rhsi},\ }\href {https://doi.org/10.1103/PhysRevResearch.2.023018} {\bibfield
  {journal} {\bibinfo  {journal} {Phys. Rev. Res.}\ }\textbf {\bibinfo {volume}
  {2}},\ \bibinfo {pages} {023018} (\bibinfo {year} {2020})}\BibitemShut
  {NoStop}%
\bibitem [{\citenamefont {Rees}\ \emph {et~al.}(2020)\citenamefont {Rees},
  \citenamefont {Manna}, \citenamefont {Lu}, \citenamefont {Morimoto},
  \citenamefont {Borrmann}, \citenamefont {Felser}, \citenamefont {Moore},
  \citenamefont {Torchinsky},\ and\ \citenamefont {Orenstein}}]{Dylan2020}%
  \BibitemOpen
  \bibfield  {author} {\bibinfo {author} {\bibfnamefont {D.}~\bibnamefont
  {Rees}}, \bibinfo {author} {\bibfnamefont {K.}~\bibnamefont {Manna}},
  \bibinfo {author} {\bibfnamefont {B.}~\bibnamefont {Lu}}, \bibinfo {author}
  {\bibfnamefont {T.}~\bibnamefont {Morimoto}}, \bibinfo {author}
  {\bibfnamefont {H.}~\bibnamefont {Borrmann}}, \bibinfo {author}
  {\bibfnamefont {C.}~\bibnamefont {Felser}}, \bibinfo {author} {\bibfnamefont
  {J.~E.}\ \bibnamefont {Moore}}, \bibinfo {author} {\bibfnamefont {D.~H.}\
  \bibnamefont {Torchinsky}},\ and\ \bibinfo {author} {\bibfnamefont
  {J.}~\bibnamefont {Orenstein}},\ }\bibfield  {title} {\bibinfo {title}
  {Helicity-dependent photocurrents in the chiral weyl semimetal rhsi},\ }\href
  {https://doi.org/10.1126/sciadv.aba0509} {\bibfield  {journal} {\bibinfo
  {journal} {Science Advances}\ }\textbf {\bibinfo {volume} {6}},\ \bibinfo
  {pages} {eaba0509} (\bibinfo {year} {2020})}\BibitemShut {NoStop}%
\bibitem [{\citenamefont {{Xu}}\ \emph {et~al.}(2020)\citenamefont {{Xu}},
  \citenamefont {{Fang}}, \citenamefont {{S{\'a}nchez-Mart{\'i}nez}},
  \citenamefont {{Venderbos}}, \citenamefont {{Ni}}, \citenamefont {{Qiu}},
  \citenamefont {{Manna}}, \citenamefont {{Wang}}, \citenamefont {{Paglione}},
  \citenamefont {{Bernhard}}, \citenamefont {{Felser}}, \citenamefont {{Mele}},
  \citenamefont {{Grushin}}, \citenamefont {{Rappe}},\ and\ \citenamefont
  {{Wu}}}]{Xu2020-zb}%
  \BibitemOpen
  \bibfield  {author} {\bibinfo {author} {\bibfnamefont {B.}~\bibnamefont
  {{Xu}}}, \bibinfo {author} {\bibfnamefont {Z.}~\bibnamefont {{Fang}}},
  \bibinfo {author} {\bibfnamefont {M.-{\'A}.}\ \bibnamefont
  {{S{\'a}nchez-Mart{\'i}nez}}}, \bibinfo {author} {\bibfnamefont {J.~W.~F.}\
  \bibnamefont {{Venderbos}}}, \bibinfo {author} {\bibfnamefont
  {Z.}~\bibnamefont {{Ni}}}, \bibinfo {author} {\bibfnamefont {T.}~\bibnamefont
  {{Qiu}}}, \bibinfo {author} {\bibfnamefont {K.}~\bibnamefont {{Manna}}},
  \bibinfo {author} {\bibfnamefont {K.}~\bibnamefont {{Wang}}}, \bibinfo
  {author} {\bibfnamefont {J.}~\bibnamefont {{Paglione}}}, \bibinfo {author}
  {\bibfnamefont {C.}~\bibnamefont {{Bernhard}}}, \bibinfo {author}
  {\bibfnamefont {C.}~\bibnamefont {{Felser}}}, \bibinfo {author}
  {\bibfnamefont {E.~J.}\ \bibnamefont {{Mele}}}, \bibinfo {author}
  {\bibfnamefont {A.~G.}\ \bibnamefont {{Grushin}}}, \bibinfo {author}
  {\bibfnamefont {A.~M.}\ \bibnamefont {{Rappe}}},\ and\ \bibinfo {author}
  {\bibfnamefont {L.}~\bibnamefont {{Wu}}},\ }\bibfield  {title} {\bibinfo
  {title} {{Optical signatures of multifold fermions in the chiral topological
  semimetal CoSi}},\ }\href {https://doi.org/10.1073/pnas.2010752117}
  {\bibfield  {journal} {\bibinfo  {journal} {Proc. Natl. Acad. Sci.}\ }\textbf
  {\bibinfo {volume} {117}},\ \bibinfo {pages} {27104} (\bibinfo {year}
  {2020})}\BibitemShut {NoStop}%
\bibitem [{\citenamefont {Le}\ \emph {et~al.}(2020)\citenamefont {Le},
  \citenamefont {Zhang}, \citenamefont {Felser},\ and\ \citenamefont
  {Sun}}]{Le2020}%
  \BibitemOpen
  \bibfield  {author} {\bibinfo {author} {\bibfnamefont {C.}~\bibnamefont
  {Le}}, \bibinfo {author} {\bibfnamefont {Y.}~\bibnamefont {Zhang}}, \bibinfo
  {author} {\bibfnamefont {C.}~\bibnamefont {Felser}},\ and\ \bibinfo {author}
  {\bibfnamefont {Y.}~\bibnamefont {Sun}},\ }\bibfield  {title} {\bibinfo
  {title} {Ab initio study of quantized circular photogalvanic effect in chiral
  multifold semimetals},\ }\href {https://doi.org/10.1103/PhysRevB.102.121111}
  {\bibfield  {journal} {\bibinfo  {journal} {Phys. Rev. B}\ }\textbf {\bibinfo
  {volume} {102}},\ \bibinfo {pages} {121111} (\bibinfo {year}
  {2020})}\BibitemShut {NoStop}%
\bibitem [{\citenamefont {Ni}\ \emph {et~al.}(2020)\citenamefont {Ni},
  \citenamefont {Xu}, \citenamefont {S{\'a}nchez-Mart{\'i}nez}, \citenamefont
  {Zhang}, \citenamefont {Manna}, \citenamefont {Bernhard}, \citenamefont
  {Venderbos}, \citenamefont {de~Juan}, \citenamefont {Felser}, \citenamefont
  {Grushin},\ and\ \citenamefont {Wu}}]{Ni2020-ze}%
  \BibitemOpen
  \bibfield  {author} {\bibinfo {author} {\bibfnamefont {Z.}~\bibnamefont
  {Ni}}, \bibinfo {author} {\bibfnamefont {B.}~\bibnamefont {Xu}}, \bibinfo
  {author} {\bibfnamefont {M.-{\'A}.}\ \bibnamefont
  {S{\'a}nchez-Mart{\'i}nez}}, \bibinfo {author} {\bibfnamefont
  {Y.}~\bibnamefont {Zhang}}, \bibinfo {author} {\bibfnamefont
  {K.}~\bibnamefont {Manna}}, \bibinfo {author} {\bibfnamefont
  {C.}~\bibnamefont {Bernhard}}, \bibinfo {author} {\bibfnamefont {J.~W.~F.}\
  \bibnamefont {Venderbos}}, \bibinfo {author} {\bibfnamefont {F.}~\bibnamefont
  {de~Juan}}, \bibinfo {author} {\bibfnamefont {C.}~\bibnamefont {Felser}},
  \bibinfo {author} {\bibfnamefont {A.~G.}\ \bibnamefont {Grushin}},\ and\
  \bibinfo {author} {\bibfnamefont {L.}~\bibnamefont {Wu}},\ }\bibfield
  {title} {\bibinfo {title} {{Linear and nonlinear optical responses in the
  chiral multifold semimetal RhSi}},\ }\href
  {https://doi.org/10.1038/s41535-020-00298-y} {\bibfield  {journal} {\bibinfo
  {journal} {npj Quantum Materials}\ }\textbf {\bibinfo {volume} {5}},\
  \bibinfo {pages} {1} (\bibinfo {year} {2020})}\BibitemShut {NoStop}%
\bibitem [{\citenamefont {Ni}\ \emph {et~al.}(2021)\citenamefont {Ni},
  \citenamefont {Wang}, \citenamefont {Zhang}, \citenamefont {Pozo},
  \citenamefont {Xu}, \citenamefont {Han}, \citenamefont {Manna}, \citenamefont
  {Paglione}, \citenamefont {Felser}, \citenamefont {Grushin}, \citenamefont
  {de~Juan}, \citenamefont {Mele},\ and\ \citenamefont {Wu}}]{Ni2021-eo}%
  \BibitemOpen
  \bibfield  {author} {\bibinfo {author} {\bibfnamefont {Z.}~\bibnamefont
  {Ni}}, \bibinfo {author} {\bibfnamefont {K.}~\bibnamefont {Wang}}, \bibinfo
  {author} {\bibfnamefont {Y.}~\bibnamefont {Zhang}}, \bibinfo {author}
  {\bibfnamefont {O.}~\bibnamefont {Pozo}}, \bibinfo {author} {\bibfnamefont
  {B.}~\bibnamefont {Xu}}, \bibinfo {author} {\bibfnamefont {X.}~\bibnamefont
  {Han}}, \bibinfo {author} {\bibfnamefont {K.}~\bibnamefont {Manna}}, \bibinfo
  {author} {\bibfnamefont {J.}~\bibnamefont {Paglione}}, \bibinfo {author}
  {\bibfnamefont {C.}~\bibnamefont {Felser}}, \bibinfo {author} {\bibfnamefont
  {A.~G.}\ \bibnamefont {Grushin}}, \bibinfo {author} {\bibfnamefont
  {F.}~\bibnamefont {de~Juan}}, \bibinfo {author} {\bibfnamefont {E.~J.}\
  \bibnamefont {Mele}},\ and\ \bibinfo {author} {\bibfnamefont
  {L.}~\bibnamefont {Wu}},\ }\bibfield  {title} {\bibinfo {title} {{Giant
  topological longitudinal circular photo-galvanic effect in the chiral
  multifold semimetal CoSi}},\ }\href
  {https://doi.org/10.1038/s41467-020-20408-5} {\bibfield  {journal} {\bibinfo
  {journal} {Nat. Commun.}\ }\textbf {\bibinfo {volume} {12}},\ \bibinfo
  {pages} {154} (\bibinfo {year} {2021})}\BibitemShut {NoStop}%
\bibitem [{\citenamefont {Rees}\ \emph {et~al.}(2021)\citenamefont {Rees},
  \citenamefont {Lu}, \citenamefont {Sun}, \citenamefont {Manna}, \citenamefont
  {\"Ozg\"ur}, \citenamefont {Subedi}, \citenamefont {Borrmann}, \citenamefont
  {Felser}, \citenamefont {Orenstein},\ and\ \citenamefont
  {Torchinsky}}]{Rees2021}%
  \BibitemOpen
  \bibfield  {author} {\bibinfo {author} {\bibfnamefont {D.}~\bibnamefont
  {Rees}}, \bibinfo {author} {\bibfnamefont {B.}~\bibnamefont {Lu}}, \bibinfo
  {author} {\bibfnamefont {Y.}~\bibnamefont {Sun}}, \bibinfo {author}
  {\bibfnamefont {K.}~\bibnamefont {Manna}}, \bibinfo {author} {\bibfnamefont
  {R.}~\bibnamefont {\"Ozg\"ur}}, \bibinfo {author} {\bibfnamefont
  {S.}~\bibnamefont {Subedi}}, \bibinfo {author} {\bibfnamefont
  {H.}~\bibnamefont {Borrmann}}, \bibinfo {author} {\bibfnamefont
  {C.}~\bibnamefont {Felser}}, \bibinfo {author} {\bibfnamefont
  {J.}~\bibnamefont {Orenstein}},\ and\ \bibinfo {author} {\bibfnamefont
  {D.~H.}\ \bibnamefont {Torchinsky}},\ }\bibfield  {title} {\bibinfo {title}
  {{Direct Measurement of Helicoid Surface States in RhSi Using Nonlinear
  Optics}},\ }\href {https://doi.org/10.1103/PhysRevLett.127.157405} {\bibfield
   {journal} {\bibinfo  {journal} {Phys. Rev. Lett.}\ }\textbf {\bibinfo
  {volume} {127}},\ \bibinfo {pages} {157405} (\bibinfo {year}
  {2021})}\BibitemShut {NoStop}%
\bibitem [{\citenamefont {Kaushik}\ and\ \citenamefont
  {Cano}(2021)}]{Kaushik2021-hx}%
  \BibitemOpen
  \bibfield  {author} {\bibinfo {author} {\bibfnamefont {S.}~\bibnamefont
  {Kaushik}}\ and\ \bibinfo {author} {\bibfnamefont {J.}~\bibnamefont {Cano}},\
  }\bibfield  {title} {\bibinfo {title} {{Magnetic photocurrents in multifold
  Weyl fermions}},\ }\href {https://doi.org/10.1103/PhysRevB.104.155149}
  {\bibfield  {journal} {\bibinfo  {journal} {Phys. Rev. B}\ }\textbf {\bibinfo
  {volume} {104}},\ \bibinfo {pages} {155149} (\bibinfo {year}
  {2021})}\BibitemShut {NoStop}%
\bibitem [{\citenamefont {Dey}\ and\ \citenamefont {Ghosh}(2022)}]{Dey2022-th}%
  \BibitemOpen
  \bibfield  {author} {\bibinfo {author} {\bibfnamefont {B.}~\bibnamefont
  {Dey}}\ and\ \bibinfo {author} {\bibfnamefont {T.~K.}\ \bibnamefont
  {Ghosh}},\ }\bibfield  {title} {\bibinfo {title} {{Dynamical polarization,
  optical conductivity and plasmon mode of a linear triple component fermionic
  system}},\ }\href {https://doi.org/10.1088/1361-648X/ac638a} {\bibfield
  {journal} {\bibinfo  {journal} {J. Phys. Condens. Matter}\ }\textbf {\bibinfo
  {volume} {34}},\ \bibinfo {pages} {255701} (\bibinfo {year}
  {2022})}\BibitemShut {NoStop}%
\bibitem [{\citenamefont {Lu}\ \emph {et~al.}(2022)\citenamefont {Lu},
  \citenamefont {Sayyad}, \citenamefont {S\'anchez-Mart\'{\i}nez},
  \citenamefont {Manna}, \citenamefont {Felser}, \citenamefont {Grushin},\ and\
  \citenamefont {Torchinsky}}]{Lu2022}%
  \BibitemOpen
  \bibfield  {author} {\bibinfo {author} {\bibfnamefont {B.}~\bibnamefont
  {Lu}}, \bibinfo {author} {\bibfnamefont {S.}~\bibnamefont {Sayyad}}, \bibinfo
  {author} {\bibfnamefont {M.~A.}\ \bibnamefont {S\'anchez-Mart\'{\i}nez}},
  \bibinfo {author} {\bibfnamefont {K.}~\bibnamefont {Manna}}, \bibinfo
  {author} {\bibfnamefont {C.}~\bibnamefont {Felser}}, \bibinfo {author}
  {\bibfnamefont {A.~G.}\ \bibnamefont {Grushin}},\ and\ \bibinfo {author}
  {\bibfnamefont {D.~H.}\ \bibnamefont {Torchinsky}},\ }\bibfield  {title}
  {\bibinfo {title} {Second-harmonic generation in the topological multifold
  semimetal rhsi},\ }\href {https://doi.org/10.1103/PhysRevResearch.4.L022022}
  {\bibfield  {journal} {\bibinfo  {journal} {Phys. Rev. Res.}\ }\textbf
  {\bibinfo {volume} {4}},\ \bibinfo {pages} {L022022} (\bibinfo {year}
  {2022})}\BibitemShut {NoStop}%
\bibitem [{\citenamefont {Pshenay-Severin}\ \emph
  {et~al.}(2018{\natexlab{b}})\citenamefont {Pshenay-Severin}, \citenamefont
  {Ivanov},\ and\ \citenamefont {Burkov}}]{Pshenay-Severin_2018_thermo}%
  \BibitemOpen
  \bibfield  {author} {\bibinfo {author} {\bibfnamefont {D.~A.}\ \bibnamefont
  {Pshenay-Severin}}, \bibinfo {author} {\bibfnamefont {Y.~V.}\ \bibnamefont
  {Ivanov}},\ and\ \bibinfo {author} {\bibfnamefont {A.~T.}\ \bibnamefont
  {Burkov}},\ }\bibfield  {title} {\bibinfo {title} {{The effect of
  energy-dependent electron scattering on thermoelectric transport in novel
  topological semimetal CoSi}},\ }\href
  {https://doi.org/10.1088/1361-648X/aae6d1} {\bibfield  {journal} {\bibinfo
  {journal} {J. Phys.: Condens. Matter}\ }\textbf {\bibinfo {volume} {30}},\
  \bibinfo {pages} {475501} (\bibinfo {year} {2018}{\natexlab{b}})}\BibitemShut
  {NoStop}%
\bibitem [{\citenamefont {Sk}\ \emph {et~al.}(2022)\citenamefont {Sk},
  \citenamefont {Shahi},\ and\ \citenamefont {Pandey}}]{Sk_2022}%
  \BibitemOpen
  \bibfield  {author} {\bibinfo {author} {\bibfnamefont {S.}~\bibnamefont
  {Sk}}, \bibinfo {author} {\bibfnamefont {N.}~\bibnamefont {Shahi}},\ and\
  \bibinfo {author} {\bibfnamefont {S.~K.}\ \bibnamefont {Pandey}},\ }\bibfield
   {title} {\bibinfo {title} {{Experimental and computational approaches to
  study the high temperature thermoelectric properties of novel topological
  semimetal CoSi}},\ }\href {https://doi.org/10.1088/1361-648X/ac655a}
  {\bibfield  {journal} {\bibinfo  {journal} {J. Phys.: Condens. Matter}\
  }\textbf {\bibinfo {volume} {34}},\ \bibinfo {pages} {265901} (\bibinfo
  {year} {2022})}\BibitemShut {NoStop}%
\bibitem [{\citenamefont {Nandy}\ \emph {et~al.}(2019)\citenamefont {Nandy},
  \citenamefont {Manna}, \citenamefont {C{\u a}lug{\u a}ru},\ and\
  \citenamefont {Roy}}]{Nandy2019-qw}%
  \BibitemOpen
  \bibfield  {author} {\bibinfo {author} {\bibfnamefont {S.}~\bibnamefont
  {Nandy}}, \bibinfo {author} {\bibfnamefont {S.}~\bibnamefont {Manna}},
  \bibinfo {author} {\bibfnamefont {D.}~\bibnamefont {C{\u a}lug{\u a}ru}},\
  and\ \bibinfo {author} {\bibfnamefont {B.}~\bibnamefont {Roy}},\ }\bibfield
  {title} {\bibinfo {title} {{Generalized triple-component fermions: Lattice
  model, Fermi arcs, and anomalous transport}},\ }\href
  {https://doi.org/10.1103/PhysRevB.100.235201} {\bibfield  {journal} {\bibinfo
   {journal} {Phys. Rev. B Condens. Matter}\ }\textbf {\bibinfo {volume}
  {100}},\ \bibinfo {pages} {235201} (\bibinfo {year} {2019})}\BibitemShut
  {NoStop}%
\bibitem [{\citenamefont {Chen}\ and\ \citenamefont
  {Wang}(2021)}]{Chen2021-dv}%
  \BibitemOpen
  \bibfield  {author} {\bibinfo {author} {\bibfnamefont {G.}~\bibnamefont
  {Chen}}\ and\ \bibinfo {author} {\bibfnamefont {C.~M.}\ \bibnamefont
  {Wang}},\ }\bibfield  {title} {\bibinfo {title} {{Optical conductivities in
  triple fermions with different monopole charges}},\ }\href
  {https://doi.org/10.1088/1361-648X/ac3d55} {\bibfield  {journal} {\bibinfo
  {journal} {J. Phys. Condens. Matter}\ }\textbf {\bibinfo {volume} {34}},\
  \bibinfo {pages} {105303} (\bibinfo {year} {2021})}\BibitemShut {NoStop}%
\bibitem [{\citenamefont {Pal}\ \emph {et~al.}(2022)\citenamefont {Pal},
  \citenamefont {Dey},\ and\ \citenamefont {Ghosh}}]{Pal2022-aw}%
  \BibitemOpen
  \bibfield  {author} {\bibinfo {author} {\bibfnamefont {O.}~\bibnamefont
  {Pal}}, \bibinfo {author} {\bibfnamefont {B.}~\bibnamefont {Dey}},\ and\
  \bibinfo {author} {\bibfnamefont {T.~K.}\ \bibnamefont {Ghosh}},\ }\bibfield
  {title} {\bibinfo {title} {{Berry curvature induced anisotropic
  magnetotransport in a quadratic triple-component fermionic system}},\ }\href
  {https://doi.org/10.1088/1361-648X/ac4cee} {\bibfield  {journal} {\bibinfo
  {journal} {J. Phys. Condens. Matter}\ }\textbf {\bibinfo {volume} {34}},\
  \bibinfo {pages} {155702} (\bibinfo {year} {2022})}\BibitemShut {NoStop}%
\bibitem [{\citenamefont {Yuan}\ \emph {et~al.}(2019)\citenamefont {Yuan},
  \citenamefont {Zhou}, \citenamefont {Rao}, \citenamefont {Tian},
  \citenamefont {Zhao}, \citenamefont {Xue}, \citenamefont {Liu}, \citenamefont
  {Zhang}, \citenamefont {Tang}, \citenamefont {Shi}, \citenamefont {Jia},
  \citenamefont {Weng}, \citenamefont {Ding}, \citenamefont {Sun},
  \citenamefont {Lei},\ and\ \citenamefont {Li}}]{Yuan2019-hz}%
  \BibitemOpen
  \bibfield  {author} {\bibinfo {author} {\bibfnamefont {Q.-Q.}\ \bibnamefont
  {Yuan}}, \bibinfo {author} {\bibfnamefont {L.}~\bibnamefont {Zhou}}, \bibinfo
  {author} {\bibfnamefont {Z.-C.}\ \bibnamefont {Rao}}, \bibinfo {author}
  {\bibfnamefont {S.}~\bibnamefont {Tian}}, \bibinfo {author} {\bibfnamefont
  {W.-M.}\ \bibnamefont {Zhao}}, \bibinfo {author} {\bibfnamefont {C.-L.}\
  \bibnamefont {Xue}}, \bibinfo {author} {\bibfnamefont {Y.}~\bibnamefont
  {Liu}}, \bibinfo {author} {\bibfnamefont {T.}~\bibnamefont {Zhang}}, \bibinfo
  {author} {\bibfnamefont {C.-Y.}\ \bibnamefont {Tang}}, \bibinfo {author}
  {\bibfnamefont {Z.-Q.}\ \bibnamefont {Shi}}, \bibinfo {author} {\bibfnamefont
  {Z.-Y.}\ \bibnamefont {Jia}}, \bibinfo {author} {\bibfnamefont
  {H.}~\bibnamefont {Weng}}, \bibinfo {author} {\bibfnamefont {H.}~\bibnamefont
  {Ding}}, \bibinfo {author} {\bibfnamefont {Y.-J.}\ \bibnamefont {Sun}},
  \bibinfo {author} {\bibfnamefont {H.}~\bibnamefont {Lei}},\ and\ \bibinfo
  {author} {\bibfnamefont {S.-C.}\ \bibnamefont {Li}},\ }\bibfield  {title}
  {\bibinfo {title} {{Quasiparticle interference evidence of the topological
  Fermi arc states in chiral fermionic semimetal CoSi}},\ }\href
  {https://doi.org/10.1126/sciadv.aaw9485} {\bibfield  {journal} {\bibinfo
  {journal} {Sci Adv}\ }\textbf {\bibinfo {volume} {5}},\ \bibinfo {pages}
  {eaaw9485} (\bibinfo {year} {2019})}\BibitemShut {NoStop}%
\bibitem [{\citenamefont {Huber}\ \emph {et~al.}(2022)\citenamefont {Huber},
  \citenamefont {Alpin}, \citenamefont {Causer}, \citenamefont {Worch},
  \citenamefont {Bauer}, \citenamefont {Benka}, \citenamefont {Hirschmann},
  \citenamefont {Schnyder}, \citenamefont {Pfleiderer},\ and\ \citenamefont
  {Wilde}}]{Huber2022}%
  \BibitemOpen
  \bibfield  {author} {\bibinfo {author} {\bibfnamefont {N.}~\bibnamefont
  {Huber}}, \bibinfo {author} {\bibfnamefont {K.}~\bibnamefont {Alpin}},
  \bibinfo {author} {\bibfnamefont {G.~L.}\ \bibnamefont {Causer}}, \bibinfo
  {author} {\bibfnamefont {L.}~\bibnamefont {Worch}}, \bibinfo {author}
  {\bibfnamefont {A.}~\bibnamefont {Bauer}}, \bibinfo {author} {\bibfnamefont
  {G.}~\bibnamefont {Benka}}, \bibinfo {author} {\bibfnamefont {M.~M.}\
  \bibnamefont {Hirschmann}}, \bibinfo {author} {\bibfnamefont {A.~P.}\
  \bibnamefont {Schnyder}}, \bibinfo {author} {\bibfnamefont {C.}~\bibnamefont
  {Pfleiderer}},\ and\ \bibinfo {author} {\bibfnamefont {M.~A.}\ \bibnamefont
  {Wilde}},\ }\bibfield  {title} {\bibinfo {title} {{Network of Topological
  Nodal Planes, Multifold Degeneracies, and Weyl Points in CoSi}},\ }\href
  {https://doi.org/10.1103/PhysRevLett.129.026401} {\bibfield  {journal}
  {\bibinfo  {journal} {Phys. Rev. Lett.}\ }\textbf {\bibinfo {volume} {129}},\
  \bibinfo {pages} {026401} (\bibinfo {year} {2022})}\BibitemShut {NoStop}%
\bibitem [{\citenamefont {Hsu}\ \emph {et~al.}(2022)\citenamefont {Hsu},
  \citenamefont {Fulga},\ and\ \citenamefont {You}}]{hsu2022disorder}%
  \BibitemOpen
  \bibfield  {author} {\bibinfo {author} {\bibfnamefont {H.-C.}\ \bibnamefont
  {Hsu}}, \bibinfo {author} {\bibfnamefont {I.~C.}\ \bibnamefont {Fulga}},\
  and\ \bibinfo {author} {\bibfnamefont {J.-S.}\ \bibnamefont {You}},\
  }\bibfield  {title} {\bibinfo {title} {Disorder effects on triple-point
  fermions},\ }\href {https://doi.org/10.1103/PhysRevB.106.245118} {\bibfield
  {journal} {\bibinfo  {journal} {Phys. Rev. B}\ }\textbf {\bibinfo {volume}
  {106}},\ \bibinfo {pages} {245118} (\bibinfo {year} {2022})}\BibitemShut
  {NoStop}%
\bibitem [{\citenamefont {Bercioux}\ \emph {et~al.}(2009)\citenamefont
  {Bercioux}, \citenamefont {Urban}, \citenamefont {Grabert},\ and\
  \citenamefont {H\"ausler}}]{Bercioux2009}%
  \BibitemOpen
  \bibfield  {author} {\bibinfo {author} {\bibfnamefont {D.}~\bibnamefont
  {Bercioux}}, \bibinfo {author} {\bibfnamefont {D.~F.}\ \bibnamefont {Urban}},
  \bibinfo {author} {\bibfnamefont {H.}~\bibnamefont {Grabert}},\ and\ \bibinfo
  {author} {\bibfnamefont {W.}~\bibnamefont {H\"ausler}},\ }\bibfield  {title}
  {\bibinfo {title} {{Massless Dirac-Weyl fermions in a ${\mathcal{T}}_{3}$
  optical lattice}},\ }\href {https://doi.org/10.1103/PhysRevA.80.063603}
  {\bibfield  {journal} {\bibinfo  {journal} {Phys. Rev. A}\ }\textbf {\bibinfo
  {volume} {80}},\ \bibinfo {pages} {063603} (\bibinfo {year}
  {2009})}\BibitemShut {NoStop}%
\bibitem [{\citenamefont {Shen}\ \emph {et~al.}(2010)\citenamefont {Shen},
  \citenamefont {Shao}, \citenamefont {Wang},\ and\ \citenamefont
  {Xing}}]{Shen2010}%
  \BibitemOpen
  \bibfield  {author} {\bibinfo {author} {\bibfnamefont {R.}~\bibnamefont
  {Shen}}, \bibinfo {author} {\bibfnamefont {L.~B.}\ \bibnamefont {Shao}},
  \bibinfo {author} {\bibfnamefont {B.}~\bibnamefont {Wang}},\ and\ \bibinfo
  {author} {\bibfnamefont {D.~Y.}\ \bibnamefont {Xing}},\ }\bibfield  {title}
  {\bibinfo {title} {Single dirac cone with a flat band touching on
  line-centered-square optical lattices},\ }\href
  {https://doi.org/10.1103/PhysRevB.81.041410} {\bibfield  {journal} {\bibinfo
  {journal} {Phys. Rev. B}\ }\textbf {\bibinfo {volume} {81}},\ \bibinfo
  {pages} {041410(R)} (\bibinfo {year} {2010})}\BibitemShut {NoStop}%
\bibitem [{\citenamefont {Vigh}\ \emph {et~al.}(2013)\citenamefont {Vigh},
  \citenamefont {Oroszl\'any}, \citenamefont {Vajna}, \citenamefont {San-Jose},
  \citenamefont {D\'avid}, \citenamefont {Cserti},\ and\ \citenamefont
  {D\'ora}}]{Vigh2013}%
  \BibitemOpen
  \bibfield  {author} {\bibinfo {author} {\bibfnamefont {M.}~\bibnamefont
  {Vigh}}, \bibinfo {author} {\bibfnamefont {L.}~\bibnamefont {Oroszl\'any}},
  \bibinfo {author} {\bibfnamefont {S.}~\bibnamefont {Vajna}}, \bibinfo
  {author} {\bibfnamefont {P.}~\bibnamefont {San-Jose}}, \bibinfo {author}
  {\bibfnamefont {G.}~\bibnamefont {D\'avid}}, \bibinfo {author} {\bibfnamefont
  {J.}~\bibnamefont {Cserti}},\ and\ \bibinfo {author} {\bibfnamefont
  {B.}~\bibnamefont {D\'ora}},\ }\bibfield  {title} {\bibinfo {title}
  {Diverging dc conductivity due to a flat band in a disordered system of
  pseudospin-1 dirac-weyl fermions},\ }\href
  {https://doi.org/10.1103/PhysRevB.88.161413} {\bibfield  {journal} {\bibinfo
  {journal} {Phys. Rev. B}\ }\textbf {\bibinfo {volume} {88}},\ \bibinfo
  {pages} {161413(R)} (\bibinfo {year} {2013})}\BibitemShut {NoStop}%
\bibitem [{\citenamefont {H{\"a}usler}(2015)}]{Hausler2015-lp}%
  \BibitemOpen
  \bibfield  {author} {\bibinfo {author} {\bibfnamefont {W.}~\bibnamefont
  {H{\"a}usler}},\ }\bibfield  {title} {\bibinfo {title} {{Flat-band
  conductivity properties at long-range Coulomb interactions}},\ }\href
  {https://doi.org/10.1103/PhysRevB.91.041102} {\bibfield  {journal} {\bibinfo
  {journal} {Phys. Rev. B}\ }\textbf {\bibinfo {volume} {91}},\ \bibinfo
  {pages} {041102(R)} (\bibinfo {year} {2015})}\BibitemShut {NoStop}%
\bibitem [{\citenamefont {Yang}\ \emph {et~al.}(2019)\citenamefont {Yang},
  \citenamefont {Chen}, \citenamefont {Shi},\ and\ \citenamefont
  {Li}}]{Yang2019-oj}%
  \BibitemOpen
  \bibfield  {author} {\bibinfo {author} {\bibfnamefont {Z.}~\bibnamefont
  {Yang}}, \bibinfo {author} {\bibfnamefont {W.}~\bibnamefont {Chen}}, \bibinfo
  {author} {\bibfnamefont {Q.~W.}\ \bibnamefont {Shi}},\ and\ \bibinfo {author}
  {\bibfnamefont {Q.}~\bibnamefont {Li}},\ }\bibfield  {title} {\bibinfo
  {title} {{Quantum conductivity correction in a two-dimensional disordered
  pseudospin-1 system}},\ }\href {https://doi.org/10.1103/PhysRevB.99.134204}
  {\bibfield  {journal} {\bibinfo  {journal} {Phys. Rev. B}\ }\textbf {\bibinfo
  {volume} {99}},\ \bibinfo {pages} {134204} (\bibinfo {year}
  {2019})}\BibitemShut {NoStop}%
\bibitem [{\citenamefont {Burgos}\ \emph {et~al.}(2022)\citenamefont {Burgos},
  \citenamefont {Warnes},\ and\ \citenamefont {Arteaga}}]{Burgos2022-rl}%
  \BibitemOpen
  \bibfield  {author} {\bibinfo {author} {\bibfnamefont {R.}~\bibnamefont
  {Burgos}}, \bibinfo {author} {\bibfnamefont {J.~H.}\ \bibnamefont {Warnes}},\
  and\ \bibinfo {author} {\bibfnamefont {G.~C.}\ \bibnamefont {Arteaga}},\
  }\bibfield  {title} {\bibinfo {title} {{Semiclassical anisotropic transport
  theory in two dimensional pseudo-spin one system}},\ }\href
  {https://doi.org/10.1016/j.physe.2021.114998} {\bibfield  {journal} {\bibinfo
   {journal} {Physica E}\ }\textbf {\bibinfo {volume} {136}},\ \bibinfo {pages}
  {114998} (\bibinfo {year} {2022})}\BibitemShut {NoStop}%
\bibitem [{\citenamefont {Ominato}\ and\ \citenamefont
  {Koshino}(2014)}]{0minato2014}%
  \BibitemOpen
  \bibfield  {author} {\bibinfo {author} {\bibfnamefont {Y.}~\bibnamefont
  {Ominato}}\ and\ \bibinfo {author} {\bibfnamefont {M.}~\bibnamefont
  {Koshino}},\ }\bibfield  {title} {\bibinfo {title} {Quantum transport in a
  three-dimensional weyl electron system},\ }\href
  {https://doi.org/10.1103/PhysRevB.89.054202} {\bibfield  {journal} {\bibinfo
  {journal} {Phys. Rev. B}\ }\textbf {\bibinfo {volume} {89}},\ \bibinfo
  {pages} {054202} (\bibinfo {year} {2014})}\BibitemShut {NoStop}%
\bibitem [{\citenamefont {Kobayashi}\ \emph {et~al.}(2014)\citenamefont
  {Kobayashi}, \citenamefont {Ohtsuki}, \citenamefont {Imura},\ and\
  \citenamefont {Herbut}}]{Kobayashi2014}%
  \BibitemOpen
  \bibfield  {author} {\bibinfo {author} {\bibfnamefont {K.}~\bibnamefont
  {Kobayashi}}, \bibinfo {author} {\bibfnamefont {T.}~\bibnamefont {Ohtsuki}},
  \bibinfo {author} {\bibfnamefont {K.-I.}\ \bibnamefont {Imura}},\ and\
  \bibinfo {author} {\bibfnamefont {I.~F.}\ \bibnamefont {Herbut}},\ }\bibfield
   {title} {\bibinfo {title} {Density of states scaling at the semimetal to
  metal transition in three dimensional topological insulators},\ }\href
  {https://doi.org/10.1103/PhysRevLett.112.016402} {\bibfield  {journal}
  {\bibinfo  {journal} {Phys. Rev. Lett.}\ }\textbf {\bibinfo {volume} {112}},\
  \bibinfo {pages} {016402} (\bibinfo {year} {2014})}\BibitemShut {NoStop}%
\bibitem [{\citenamefont {Nandkishore}\ \emph {et~al.}(2014)\citenamefont
  {Nandkishore}, \citenamefont {Huse},\ and\ \citenamefont
  {Sondhi}}]{Nandkishore2014-vz}%
  \BibitemOpen
  \bibfield  {author} {\bibinfo {author} {\bibfnamefont {R.}~\bibnamefont
  {Nandkishore}}, \bibinfo {author} {\bibfnamefont {D.~A.}\ \bibnamefont
  {Huse}},\ and\ \bibinfo {author} {\bibfnamefont {S.~L.}\ \bibnamefont
  {Sondhi}},\ }\bibfield  {title} {\bibinfo {title} {{Rare region effects
  dominate weakly disordered three-dimensional Dirac points}},\ }\href
  {https://doi.org/10.1103/PhysRevB.89.245110} {\bibfield  {journal} {\bibinfo
  {journal} {Phys. Rev. B}\ }\textbf {\bibinfo {volume} {89}},\ \bibinfo
  {pages} {245110} (\bibinfo {year} {2014})}\BibitemShut {NoStop}%
\bibitem [{\citenamefont {Ominato}\ and\ \citenamefont
  {Koshino}(2015)}]{Ominato2015-um}%
  \BibitemOpen
  \bibfield  {author} {\bibinfo {author} {\bibfnamefont {Y.}~\bibnamefont
  {Ominato}}\ and\ \bibinfo {author} {\bibfnamefont {M.}~\bibnamefont
  {Koshino}},\ }\bibfield  {title} {\bibinfo {title} {Quantum transport in
  three-dimensional weyl electron system in the presence of charged impurity
  scattering},\ }\href {https://doi.org/10.1103/PhysRevB.91.035202} {\bibfield
  {journal} {\bibinfo  {journal} {Phys. Rev. B}\ }\textbf {\bibinfo {volume}
  {91}},\ \bibinfo {pages} {035202} (\bibinfo {year} {2015})}\BibitemShut
  {NoStop}%
\bibitem [{\citenamefont {Ominato}\ and\ \citenamefont
  {Koshino}(2016)}]{Ominato2016-jl}%
  \BibitemOpen
  \bibfield  {author} {\bibinfo {author} {\bibfnamefont {Y.}~\bibnamefont
  {Ominato}}\ and\ \bibinfo {author} {\bibfnamefont {M.}~\bibnamefont
  {Koshino}},\ }\bibfield  {title} {\bibinfo {title} {{Magnetotransport in Weyl
  semimetals in the quantum limit: Role of topological surface states}},\
  }\href {https://doi.org/10.1103/PhysRevB.93.245304} {\bibfield  {journal}
  {\bibinfo  {journal} {Phys. Rev. B}\ }\textbf {\bibinfo {volume} {93}},\
  \bibinfo {pages} {245304} (\bibinfo {year} {2016})}\BibitemShut {NoStop}%
\bibitem [{\citenamefont {Noro}\ \emph {et~al.}(2010)\citenamefont {Noro},
  \citenamefont {Koshino},\ and\ \citenamefont {Ando}}]{Noro2010-cm}%
  \BibitemOpen
  \bibfield  {author} {\bibinfo {author} {\bibfnamefont {M.}~\bibnamefont
  {Noro}}, \bibinfo {author} {\bibfnamefont {M.}~\bibnamefont {Koshino}},\ and\
  \bibinfo {author} {\bibfnamefont {T.}~\bibnamefont {Ando}},\ }\bibfield
  {title} {\bibinfo {title} {{Theory of Transport in Graphene with Long-Range
  Scatterers}},\ }\href {https://doi.org/10.1143/JPSJ.79.094713} {\bibfield
  {journal} {\bibinfo  {journal} {J. Phys. Soc. Jpn.}\ }\textbf {\bibinfo
  {volume} {79}},\ \bibinfo {pages} {094713} (\bibinfo {year}
  {2010})}\BibitemShut {NoStop}%
\bibitem [{\citenamefont {Jay-Gerin}\ \emph {et~al.}(1977)\citenamefont
  {Jay-Gerin}, \citenamefont {Aubin},\ and\ \citenamefont
  {Caron}}]{J.-P.Jay-Gerin1977}%
  \BibitemOpen
  \bibfield  {author} {\bibinfo {author} {\bibfnamefont {J.-P.}\ \bibnamefont
  {Jay-Gerin}}, \bibinfo {author} {\bibfnamefont {M.}~\bibnamefont {Aubin}},\
  and\ \bibinfo {author} {\bibfnamefont {L.}~\bibnamefont {Caron}},\ }\bibfield
   {title} {\bibinfo {title} {The electron mobility and the static dielectric
  constant of cd3as2 at 4.2 k},\ }\href
  {https://doi.org/https://doi.org/10.1016/0038-1098(77)91149-8} {\bibfield
  {journal} {\bibinfo  {journal} {Solid State Communications}\ }\textbf
  {\bibinfo {volume} {21}},\ \bibinfo {pages} {771} (\bibinfo {year}
  {1977})}\BibitemShut {NoStop}%
\end{thebibliography}%
\clearpage
\end{document}